\documentclass[%
reprint,
 amsmath,amssymb,
 aps,
prd,
]{revtex4-2}

\usepackage{graphicx}
\usepackage{dcolumn}
\usepackage{bm}

\usepackage{fancyhdr}
\usepackage{amsmath}
\usepackage{amssymb}
\usepackage{booktabs}
\usepackage{graphicx}
\usepackage{caption}
\usepackage[a4paper, margin=1in]{geometry}
\usepackage{graphicx} 
\usepackage{xcolor}

\usepackage[compat=1.1.0]{tikz-feynman}
\tikzfeynmanset{every diagram/.style={small}}
\usepackage{float}
\usepackage{booktabs}
\usepackage[colorlinks=true,linktocpage=true,linkcolor=blue,citecolor=teal]{hyperref}
\usepackage[normalem]{ulem}
\usepackage{siunitx} 
\usepackage{caption}
\usepackage{tikz}
\usepackage{tikz-feynman}
\tikzfeynmanset{compat=1.1.0} 

\usetikzlibrary{positioning}
\usepackage[utf8]{inputenc}  
\DeclareUnicodeCharacter{202F}{} 

\begin{document}


\title{Prospects for a fourth generation of leptons in a 13 TeV $p-p$ collider \\}

\author{Ramkrishna Joshi}
\email{joram0911@gmail.com}
\author{Riddhiman Roy}%
 \email{riddhiman.roy1911@proton.me}
\affiliation{%
 Department of Physics, Indian Institute of Technology Hyderabad, Telangana, Hyderabad\\
}%

\date{\today}

\begin{abstract}

   In the Standard Model, three discovered generations of leptons and quarks are known to date. However, speculations about existence of next generations have a strong foothold. In this study, we sequentially extrapolate the Standard Model to include a fourth generation of leptons $(\ell_4,\nu_4)$ with a massive Dirac neutrino. We perform MC simulated event generation of  $pp \rightarrow \ell_4 \overline{\ell_4}$ scattering processes at an LHC-like $p-p$ collider with $\sqrt{s} = 13\ $TeV by considering $\ell_{4}$  mass of $190\ $ GeV and $\nu_{4}$ mass of $100\ $GeV with PYTHIA. We demonstrate mass constraints of sequential leptons from oblique parameters and study important jet and lepton kinematics in our simulation with CMS like constraints. Fourth generation neutrino is stable in this scenario making $\ell_4 \rightarrow W \nu_{4}$ the only dominant channel in collider searches.With the cut-flow, we achieve global excess of $(1.46 \pm0.068 (stat.))\sigma$ and local excess of $(3.33 \pm 0.241(stat.))\sigma$ in the $180-300\ GeV$ signal window. Missing Transverse Energy ($\text{ME}_T$) provides clean signature of $\nu_4$. We assess discovery potential of the BSM lepton against lepton mass,center-of-mass energy (E$_{com}$) and luminosities. Higher luminosities and E$_{com}$ are promising to probe BSM moderate mass lepton scenarios at present and future colliders.\\

   \textbf{Keywords:} BSM lepton sector, heavy dirac neutrino, oblique parameters, jet kinematics

\end{abstract}

\maketitle


\section{Introduction\protect\\}




 The existence of higher generations of quarks and leptons open horizons for new Beyond Standard Model (BSM) physics. One of the most distinct BSM particles is the neutrino, which depends significantly on the $Z$ boson mass and decay width measurements. Electron-positron colliders such as LEP and SLC have operated at the $Z$ resonance and provide highly precise measurements of total decay width $\Gamma_Z$, which can be used to constrain the number of light neutrino generations \cite{ALEPH_1989,DELPHI_1989,OPAL_1989, MarkII_1989a, MarkII_1989b}.

The partial width for the decay of the $Z$ boson into a single neutrino--anti-neutrino pair is given by \cite{2023137563}:
\begin{equation}
\Gamma(Z \to \nu \bar{\nu}) = \frac{G_F M_Z^3}{12 \pi \sqrt{2}} \left( g_V^2 + g_A^2 \right),
\end{equation}

where $G_F$ is the Fermi coupling constant, $M_Z$ is the $Z$ boson mass, and $g_V$, $g_A$ are the vector and axial-vector couplings of the neutrino to the $Z$. For neutrinos, $g_V = g_A = \frac{1}{2}$ and hence partial width and total invisible width is obtained by,
\begin{equation}
\Gamma(Z \to \nu \bar{\nu}) = \frac{G_F M_Z^3}{24 \pi \sqrt{2}}, \ \Gamma_{\text{inv}} = \Gamma_Z - \sum_{f \neq \nu} \Gamma(Z \to f \bar{f}),
\end{equation}

To reduce systematic uncertainties, the ratio
\begin{equation}
R_{\text{inv}} = N_\nu \cdot \frac{\Gamma(Z \to \nu \bar{\nu})}{\Gamma_{\ell\ell}}.
\end{equation}

is formed, where $\Gamma_{\ell\ell}$ is the partial width into a single charged lepton pair (e.g., $e^+e^-$) \cite{2023137563, PDG2017_ZBoson}. This ratio is proportional to the number of light neutrino species $N_\nu$. Using the Standard Model prediction for the partial widths, the LEP measurements showed the value of $N_\nu$ to be $2.984 \pm 0.008$ \cite{DELPHI_1989, OPAL_1989,ALEPH_1989, PDG1996}. The measured value from experimental results agrees very well with the Standard Model prediction of three light neutrino generations \cite{ParticleDataGroup:2024cfk}. A fourth neutrino with mass less than $M_Z/2 \sim 45$ GeV would increase $\Gamma_{\text{inv}}$ by approximately 167 MeV, leading to a clear deviation from the experimental measurement. However, the existence of a fourth neutrino with mass above $M_Z/2$, or a sterile neutrino that does not couple to the $Z$ boson, is not excluded by this result. In this work, we extend the Standard Model (SM) to include a sequential fourth generation of leptons which include a charged lepton \( \ell_4 \), and a corresponding neutrino \( \nu_4 \), both treated as Dirac particles. The $W$ and $Z$ boson couplings of $\ell_4$ make it prone to LEP bounds \cite{Rajaraman_2010:a, Rajaraman_2010:b} and electroweak precision constraints. This work is presented in the following manner: Section (2) includes technical aspects of model initialization in  SARAH and coupling vertices for fourth generation lepton. Section (3) describes constraints on masses of leptons from EWPO and LEP constraints. We provide a detailed account of contribution of large mass splittings to EWPO and comment on their consistency with respect to our model. In section (4) and (5), we provide cross section of the concerned production diagram at different center of mass energies and mass hypotheses for $\ell_4$ and we perform comparison of this cross section with $t\bar{t},WW,ZZ$ backgrounds. In section (5) provide cutflow and selection efficiencies for signal and background. Section (6) encompasses the collider signatures and jet,lepton kinematics. The concluding section (7) provides comments on discovery significance of the BSM lepton as a function of parameters mass,center of mass energy and luminosity. In the conclusion we provide comments on how can these signatures be probed at present and future colliders and distinguishing features of the signal.

\section{Fourth generation lepton model}

The extension of the SM to include a fourth generation sequential leptons is carried out using SARAH 4.15.3 \cite{Staub_2014}. The Yukawa sector of the model accounts for lepton mass generation via the Higgs mechanism. Before Electroweak Symmetry Breaking (EWSB), the lepton Yukawa interactions in the gauge basis are given by:
\begin{align}
\mathcal{L}_{\text{Yuk}}^{\text{leptons}} &\supset 
 - H^0\, \bar{e}_{L,k}\, Y_{e,jk}^{*}\, e_{R,j} \\ \nonumber
 & - H^+\, \bar{\nu}_{L,k}\, Y_{e,jk}^{*}\, e_{R,j} 
 + \text{h.c.},
\end{align}
where \( H^0 \) and \( H^+ \) are the neutral and charged components of the Higgs doublet, and \( Y_e \) is the charged lepton Yukawa matrix \cite{Tong_StandardModel6, Gesteau2021}. The indices \( j, k \in \{1,2,3,4\} \) account for the four generations.

Upon spontaneous symmetry breaking, the Higgs doublet acquires a Vacuum Expectation Value (VEV), \( \langle H^0 \rangle = v/\sqrt{2} \), which leads to mass terms for the charged leptons and neutrinos \cite{KeusKingMoretti2013}. The Dirac mass for the charged lepton is then expressed as

\begin{equation}
m_e = \frac{v}{\sqrt{2}} Y_e, m_\nu = \frac{v}{\sqrt{2}} Y_\nu
\end{equation}

where \( Y_\nu \) is the neutrino Yukawa matrix \cite{Tong_StandardModel6}. We consider the model with fourth generation neutrino of mass 100 GeV and lepton mass of 190 GeV, which will be discussed further in the following section. We maintain the mass splitting of 90 GeV between $\nu_4$ and $\ell_4$. For the fourth generation neutrino to not contribute to the decay width of $Z$ boson, the mass constraint on neutrino requires the neutrino mass to be greater than half the mass of $Z$ boson. Hence chosen mass falls within acceptable range.

Gauge interactions of the fourth-generation leptons follow from the standard electroweak gauge structure \cite{SpringerChap_978-3-662-03841-3_11}. We explicitly set the fourth-generation neutrino \( \nu_4 \) to be stable. We consider that the two potential reasons for its stability are either a symmetry (e.g., a conserved \( \mathbb{Z}_2 \)) or the suppression of mixing with lighter SM neutrinos. This nature of \( \nu_4 \) allows for its signature in our analysis as missing transverse momentum. From a collider physics perspective, the decay \( \ell_4^- \rightarrow W^- + \nu_4 \) leads to signatures involving a $W^-$ boson (decaying hadronically or leptonically) and invisible final states, offering clean experimental channels for searches at current or future colliders.

The electroweak interaction vertices involving the fourth-generation leptons emerge naturally from the SU(2)$_L$ $\times$ U(1)$_Y$, extended to include a sequential fourth generation. In this framework, the fourth-generation charged lepton, $\ell_4$, and the corresponding neutrino, $\nu_4$, acquire gauge and Yukawa interactions that are similar to their existing counterparts ($e, \mu, \tau$). However, the phenomenology of these searches will be different due to their heaviness. 

We present below the key electroweak interactions of $\ell_4$ and $\nu_4$ with the $W$, $Z$, and Higgs bosons, derived from the SARAH-generated Lagrangian in the EWSB basis. 

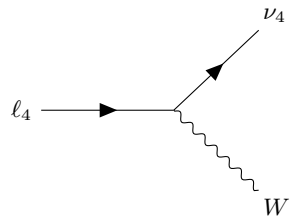
\begin{figure}[H]
\centering
\begin{tikzpicture}
\begin{feynman}
\vertex (a) {$\ell_4$};
\vertex [right=2cm of a] (b);
\vertex [above right=1.5cm of b] (f1) {$\nu_4$};
\vertex [below right=1.5cm of b] (f2) {$W$};
\diagram* {
(a) -- [fermion] (b) -- [fermion] (f1),
(b) -- [boson] (f2),
};
\end{feynman}
\end{tikzpicture}
\caption{W (Charged) vertex}
\label{fig:cc}
\end{figure}

The charged current interaction is mediated by the $W$ boson and connects $\ell_4$ and $\nu_4$ via the standard V--A structure.
\begin{equation}
\mathcal{L}_{W} = -\frac{g}{\sqrt{2}} \bar{\nu}_4 \gamma^\mu P_L \ell_4 W^+_\mu + \text{h.c.}
\end{equation}
and the corresponding Feynman diagram is shown in Figure~\ref{fig:cc}.

In non charge interactions, both $\ell_4$ and $\nu_4$ couple to the $Z$ boson. The neutrino couples via a purely left-handed vertex, while the charged lepton interaction includes both vector and axial-vector components due to its electric charge and weak isospin. The Lagrangian terms are expressed as
\begin{align}
\mathcal{L}_{Z,\nu_4} &= -\frac{g}{2\cos\theta_W} \bar{\nu}_4 \gamma^\mu P_L \nu_4 Z_\mu \\
\mathcal{L}_{Z,\ell_4} &= -\frac{g}{2\cos\theta_W} \bar{\ell}_4 \gamma^\mu \left( -P_L + 2\sin^2\theta_W \right) \ell_4 Z_\mu.
\end{align}

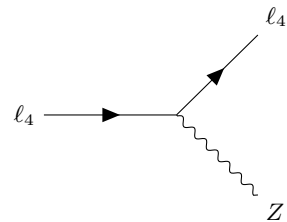
\begin{figure}[H]
\centering
\begin{tikzpicture}
\begin{feynman}
\vertex (a) {$\ell_4$};
\vertex [right=2cm of a] (b);
\vertex [above right=1.5cm of b] (f1) {$\ell_4$};
\vertex [below right=1.5cm of b] (f2) {$Z$};
\diagram* {
(a) -- [fermion] (b) -- [fermion] (f1),
(b) -- [boson] (f2),
};
\end{feynman}
\end{tikzpicture}
\caption{$Z$ (Neutral) vertex}
\label{fig:nc}
\end{figure}

Electromagnetic interactions of $\ell_4$ proceed through the standard QED vertex with the photon. Since $\nu_4$ is electrically neutral, it does not couple directly to the photon. The QED Lagrangian for $\ell_4$ is

\begin{equation}
\mathcal{L}_{\gamma} = - e \bar{\ell}_4 \gamma^\mu \ell_4 A_\mu.
\end{equation}

\begin{figure}[H]
\centering
\begin{tikzpicture}
\begin{feynman}
\vertex (a) {$\ell_4$};
\vertex [right=2cm of a] (b);
\vertex [above right=1.5cm of b] (f1) {$\ell_4$};
\vertex [below right=1.5cm of b] (f2) {$\gamma$};
\diagram* {
(a) -- [fermion] (b) -- [fermion] (f1),
(b) -- [boson] (f2),
};
\end{feynman}
\end{tikzpicture}
\caption{$\gamma$ (EM) vertex}
\label{fig:em}
\end{figure}
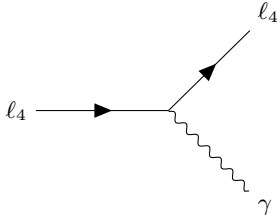

The Yukawa couplings of the fourth-generation leptons to the Higgs boson are responsible for their masses and yield scalar interaction vertices with the physical Higgs field $h$. These couplings take the form
\begin{align}
\mathcal{L}_{h,\ell_4} &= -\frac{i}{\sqrt{2}} \bar{\ell}_4 \left( Y_{\ell_4} P_R + Y_{\ell_4}^* P_L \right) \ell_4 h \\
\mathcal{L}_{h,\nu_4} &= -\frac{i}{\sqrt{2}} \bar{\nu}_4 \left( Y_{\nu_4} P_R + Y_{\nu_4}^* P_L \right) \nu_4 h.
\end{align}

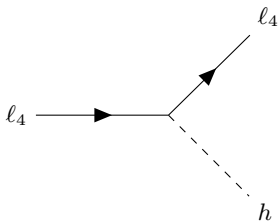
\begin{figure}[H]
\centering
\begin{tikzpicture}
\begin{feynman}
\vertex (a) {$\ell_4$};
\vertex [right=2cm of a] (b);
\vertex [above right=1.5cm of b] (f1) {$\ell_4$};
\vertex [below right=1.5cm of b] (f2) {$h$};
\diagram* {
(a) -- [fermion] (b) -- [fermion] (f1),
(b) -- [scalar] (f2),
};
\end{feynman}
\end{tikzpicture}
\caption{Higgs Yukawa coupling to $\ell_4$}
\label{fig:higgs}
\end{figure}

Finally, in extended Higgs sectors where a charged scalar $H^\pm$ exists, additional interactions involving $\ell_4$ and $\nu_4$ are possible. The coupling to the charged Higgs is typically governed by the same Yukawa structure, leading to the interaction
\begin{equation}
\mathcal{L}_{H^\pm} = -i Y_{\ell_4} \bar{\nu}_4 P_R \ell_4 H^+ + \text{h.c.}
\end{equation}

To diagonalize the mass matrices of the fourth-generation leptons, unitary rotations are applied to the gauge basis states. For the charged lepton sector, the mass matrix $m_e$ is diagonalized via:
\begin{equation}
U_{eL}^\dagger m_e U_{eR} = \text{diag}(m_e, m_\mu, m_\tau, m_{\ell_4}).
\end{equation}
The left-handed rotation matrix $U_{eL}$ also appears in the charged current interaction vertex, governing the coupling of $\ell_4$ to $W$ and $\nu_4$. While mixing with the first three generations allow for possibility of Lepton Flavor Violating (LFV) processes, we assume such mixing to be negligible for convenience of the model and to relax some constraint induced by such processes. This assumption relaxes tight constraints from LFV decay searches and electroweak precision observables. CP-odd Higgs interactions and exotic decays are also potentially observable. The scalar sector of the model includes a CP-odd neutral scalar $A^0$, which couples to the fourth-generation leptons. The interaction Lagrangian has terms of the form:
\begin{equation}
\mathcal{L}_{A^0} \supset \bar{e}_i i\gamma^5 Y_{e,ij} e_j A^0 + \text{h.c.}
\end{equation}
This allows processes such as $\ell_4 \to \ell_i A^0$, if they are permitted kinematically. These give rise to hard photons due to very heavy $\ell_4$.

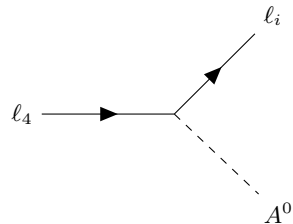
\begin{figure}[H]
\centering
\begin{tikzpicture}
\begin{feynman}
\vertex (a) {$\ell_4$};
\vertex [right=2cm of a] (b);
\vertex [above right=1.5cm of b] (f1) {$\ell_i$};
\vertex [below right=1.5cm of b] (f2) {$A^0$};
\diagram* {
(a) -- [fermion] (b) -- [fermion] (f1),
(b) -- [scalar] (f2),
};
\end{feynman}
\end{tikzpicture}
\caption{Decay $\ell_4 \to \ell_i A^0$ via pseudoscalar Yukawa coupling}
\label{fig:l4_to_A0}
\end{figure}

If $A^0$ is sufficiently light, it may decay to invisible final states such as $\nu_4 \bar{\nu}_4$:

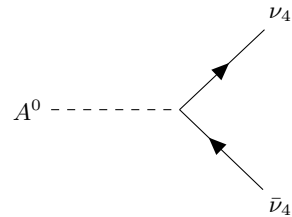
\begin{figure}[H]
\centering
\begin{tikzpicture}
\begin{feynman}
\vertex (a) {$A^0$};
\vertex [right=2cm of a] (b);
\vertex [above right=1.5cm of b] (f1) {$\nu_4$};
\vertex [below right=1.5cm of b] (f2) {$\bar{\nu}_4$};
\diagram* {
(a) -- [scalar] (b),
(b) -- [fermion] (f1),
(b) -- [anti fermion] (f2),
};
\end{feynman}
\end{tikzpicture}
\caption{Invisible decay $A^0 \to \nu_4 \bar{\nu}_4$}
\label{fig:A0_nu}
\end{figure}

To preserve gauge invariance after quantization, the model implements an $R_\xi$ gauge-fixing procedure. The gauge-fixing Lagrangian introduces ghost fields corresponding to each gauge boson, including $\eta^\pm$, $\eta^Z$, and $\eta^\gamma$. SARAH automatically generates the associated ghost interactions, extended to include the fourth-generation lepton content. However, these are not directly associated with the phenomenology of this work and hence we leave it to the reader to study more about them.

A key collider signature is the decay chain:
\[
\ell_4^- \to W^- \nu_4, \quad W^- \to \ell^- \bar{\nu}
\]
and its counterpart which results in a dilepton plus missing energy final state.

\begin{figure}[H]
\centering
\begin{tikzpicture}
\begin{feynman}
\vertex (a) {$\ell_4^-$};
\vertex [right=2cm of a] (b);
\vertex [above right=1.5cm of b] (f1) {$\nu_4$};
\vertex [below right=1.5cm of b] (c);
\vertex [above right=1.5cm of c] (f2) {$\bar{\nu/\bar{q}}$};
\vertex [below right=1.5cm of c] (f3) {$\ell^-/q$};
\diagram* {
(a) -- [fermion] (b) -- [fermion] (f1),
(b) -- [boson, edge label'=$W^-$] (c),
(c) -- [anti fermion] (f2),
(c) -- [fermion] (f3),
};
\end{feynman}
\end{tikzpicture}
\caption{Cascade decay $\ell_4^- \to W^- \nu_4 \to \ell^- \bar{\nu} \nu_4$}
\label{fig:l4_decay}
\end{figure}
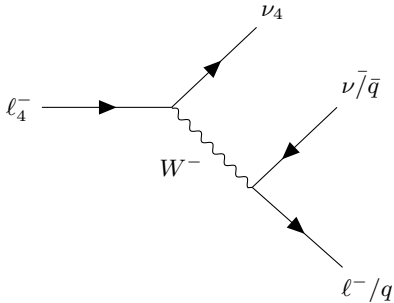

In conclusion, the fourth-generation leptons interaction vertices are identical in structure to their lighter SM counterparts, but the phenomenology can significantly differ due to mentioned reasons.

Considering that the dominant decay mode of the fourth generation lepton is to $W$ boson and a heavy fourth generation neutrino, the partial decay width for this process, including the effect of a non-negligible neutrino mass, is given by \cite{Atre2009}:

\begin{align}
\Gamma(\ell_4 \to W \nu_4) &=
\frac{g^2}{64 \pi} \,
\frac{m_{\ell_4}^3}{m_W^2} \,
\lambda^{1/2}\nonumber
\times \\&\left[
\left(1 - \frac{m_W^2 + m_{\nu_4}^2}{m_{\ell_4}^2}\right)^2
- \frac{4 m_W^2 m_{\nu_4}^2}{m_{\ell_4}^4}
\right]^{1/2}\nonumber
\\&\times \left(
1 + \frac{m_W^2 - m_{\nu_4}^2}{m_{\ell_4}^2}
\right),
\end{align}

with
$\lambda =
\left(1 - \frac{(m_W + m_{\nu_4})^2}{m_{\ell_4}^2}\right)
\left(1 - \frac{(m_W - m_{\nu_4})^2}{m_{\ell_4}^2}\right).$\\

Here, $g$ is the SU(2)$_L$ weak coupling constant and m are masses. This fully accounts for the phase space suppression from both final state masses. Using $g = 0.653$, we calculate decay widths for different combinations of lepton masses. The results are summarized in Table~(\ref{tab:l4widths_updated}). For a fixed mass of $\nu_4$, decay width scales rapidly, as expected.

\begin{table}[H]
\centering
\small
\setlength{\tabcolsep}{3pt} 
\caption{Partial decay widths for two benchmark scenarios:
(i) fixed $m_{\nu_4}=100$ GeV, and
(ii) fixed mass splitting $\Delta m = m_{\ell_4}-m_{\nu_4}=90$ GeV.}
\label{tab:l4widths_updated}
\begin{tabular}{c c c c c}
\hline\hline
$m_{\ell_4}$ & $m_{\nu_4}$ (fixed) & $\Gamma$ & $m_{\nu_4}$ ($\Delta m=90$) & $\Gamma$ \\
(GeV) & (GeV) & (GeV) & (GeV) & (GeV) \\
\hline
190  & 100 & 0.198 & 100 & 0.198 \\
500  & 100 & 35.1  & 410 & 0.313 \\
750  & 100 & 129.6 & 660 & 0.337 \\
1000 & 100 & 316.3 & 910 & 0.349 \\
\hline\hline
\end{tabular}
\end{table}

These decay widths are considered while reconstructing the $\ell_4$ from the decay products of the $W$ boson and the stable $\nu_4$, ensuring realistic modeling of invariant mass distributions in the analysis. In this model, the matter content is extended to include a fourth generation of leptons. The table below summarizes the superfields defined in the model, along with their spin, number of generations, and gauge quantum numbers under the Standard Model gauge group \( U(1)_Y \otimes SU(2)_L \otimes SU(3)_C \). After EWSB, the physical spectrum includes the mass eigenstates of the fields. The table below lists all particle content, indicating their type, real or complex nature, number of generations, and index structure.

\begin{table}[H]
\centering
\small
\setlength{\tabcolsep}{4pt}
\caption{Matter superfields in the EWSB basis.}
\label{tab:matter_superfields}
\resizebox{\columnwidth}{!}{
\begin{tabular}{c c c c c}
\hline\hline
Name & Spin/Type & Generations & $U(1)_Y \otimes SU(2)_L \otimes SU(3)_C$ \\
\hline
$H$ & Scalar (complex) & 1 & $\left( \tfrac{1}{2}, \mathbf{2}, \mathbf{1} \right)$ \\
$q$ & Fermion & 3 & $\left( \tfrac{1}{6}, \mathbf{2}, \mathbf{3} \right)$ \\
$l$ & Fermion & 4 & $\left( -\tfrac{1}{2}, \mathbf{2}, \mathbf{1} \right)$ \\
$d$ & Fermion & 3 & $\left( \tfrac{1}{3}, \mathbf{1}, \bar{\mathbf{3}} \right)$\\
$u$ & Fermion & 3 & $\left( -\tfrac{2}{3}, \mathbf{1}, \bar{\mathbf{3}} \right)$ \\
$e$ & Fermion & 4 & $\left( 1, \mathbf{1}, \mathbf{1} \right)$ \\
\hline\hline
\end{tabular}
}
\end{table}

With matter superfields we can represent the particle contents of the model. The sequential extension of the standard model can be readily interpreted from the generations column for matter superfields. The same reflects in particle contents also as seen from Table (\ref{tab:ewsb_spectrum}) below. Both BSM leptons are dirac.

\begin{table}[H]
\centering
\small
\caption{Particle content after EWSB.}
\label{tab:ewsb_spectrum}
\resizebox{\columnwidth}{!}{%
\begin{tabular}{c c c c c}
\hline\hline
Name & Type & Generations & Real/Complex & Indices \\
\hline
$H^+$, $A^0$, $h$ & Scalars & 1 & real/complex & -- \\
$\nu$ & Fermion & 4 & Dirac & generation \\
$d$, $u$ & Fermion & 3 & Dirac & gen.\ 3, color 3 \\
$e$ & Fermion & 4 & Dirac & generation \\
$g$ & Vector & 1 & real & color 8, Lorentz 4 \\
$\gamma$, $Z$ & Vector & 1 & real & Lorentz 4 \\
$W^\pm$ & Vector & 1 & complex & Lorentz 4 \\
$\eta^G$ & Ghost & 1 & real & color 8 \\
$\eta^\gamma$, $\eta^Z$ & Ghost & 1 & real & -- \\
$\eta^\pm$ & Ghost & 1 & complex & -- \\
\hline\hline
\end{tabular}%
}
\end{table}

We study this  phenomenology for pp collisions at $\sqrt{s} = 13$ TeV E$_{com}$ with symmetric proton beams. In following sections, we provide detailed analysis of constraints on fourth generation fermion masses followed by analysis of jet $p_T$, jet multiplicity, missing transverse momentum, invariant reconstructed mass, dijet invariant mass, lepton $p_T$ and lepton multiplicity for $(m_{E_4},m_{N_4}) = (190 \ \text{GeV}, 100 \ \text{GeV})$. We also provide detailed PYTHIA analysis with justification of cut flow, signal selection efficiency, background suppression techniques used and Initial State Radiation (ISR)/ Final State Radiation (FSR) effects on parameters of interest. For the BSM neutrino, mass value is chosen from available constraints. For the lepton the reference value of $190\ GeV$ is chosen to account for minimum threshold $m_{\ell_4}=m_W+m_{\nu{4}}$ which will kinematically allow relevant diagrams. Thus, our analysis provides a collider signature framework for threshold masses of sequential pair in BSM lepton sector.

\section{Constraints on masses of sequential leptons}


The choice of a fourth-generation charged lepton mass of 190 GeV and a neutrino mass of 100 GeV is consistent with existing experimental and theoretical constraints. Direct collider searches place the following lower bounds on these masses \cite{Nakamura_2010}:

\begin{equation}
m_E > 100.8~\text{GeV},~ m_N > (80.5 - 101.5)~\text{GeV}
\end{equation}

where the range for $m_N$ arises from different LEP search channels assuming a $100\%$ branching ratio to $W^* \ell$ in each channel \cite{Rajaraman_2010:b}. These limits are slightly weaker for Majorana neutrinos compared to Dirac neutrinos by approximately 10 GeV. Electroweak precision tests impose additional constraints on the mass splitting between the fourth-generation charged lepton ($E$) and neutrino ($N$), expressed as \cite{PhysRevD.82.095006}:

\begin{equation}
|m_E - m_N| < 140~\text{GeV},
\end{equation}

to avoid excessive contributions to the oblique parameter $T$ \cite{PhysRevD.46.381}. In the benchmark considered here, with $m_{\ell_4} = 190$ GeV and $m_{\nu_4} = 100$ GeV, the mass splitting is 90 GeV, satisfying this requirement. Furthermore, the PMNS matrix elements involving the fourth generation are constrained by rare decay processes. The upper limits on mixing angles and an additional strong bound from $\mu - e$ conversion for $m_N > 100$ GeV are \cite{Deshpande2011, Nakamura_2010,PhysRevD.46.381};

\begin{align}
|U_{e4}|   &< 0.073, 
& |U_{\mu4}|   &< 0.045, \notag \\
|U_{\tau4}|&< 0.072, 
& |U_{E4}|     &> 0.9958, \notag \\
|U_{\mu4}^* U_{e4}| &< 0.4 \times 10^{-4} &&
\end{align}


The contributions of a sequential fourth-generation lepton doublet to the oblique parameters are significant, particularly for the $T$ parameter, which increases with mass splitting $\Delta m = |m_{\ell_4} - m_{\nu_4}|$ due to custodial symmetry breaking. Here we have chosen the mass values of lepton and neutrino to be such that we work in the limit $m_{\nu_4},m_{\ell_4}>m_Z$ and hence the contribution to $T$ is approximately \cite{PhysRevD.46.381}:

\begin{align}
\Delta T &\approx \frac{N_c}{16\pi s_W^2 c_W^2 m_Z^2}\times\\\nonumber
&\left[
m_{\ell_4}^2 + m_{\nu_4}^2
- \frac{2 m_{\ell_4}^2 m_{\nu_4}^2}{m_{\ell_4}^2 - m_{\nu_4}^2}
\ln \left( \frac{m_{\ell_4}^2}{m_{\nu_4}^2} \right)
\right],
\end{align}

where $N_c=1$ for leptons, and $s_W$, $c_W$ are the sine and cosine of the weak mixing angle. For $m_{\ell_4} = 190$ GeV and $m_{\nu_4} = 100$ GeV, using $s_W^2 = 0.231$ \cite{ParticleDataGroup:2024cfk}, $c_W^2 = 0.769$ \cite{ParticleDataGroup:2024cfk}, and $m_Z = 91.1876$ GeV, we get $\Delta T \approx 0.143$. In the same mass limit, the S parameter contributes as,
\begin{equation}
\Delta S \approx \frac{1}{6\pi}
\left[
1 - Y \ln \left( \frac{m_{\ell_4}^2}{m_{\nu_4}^2} \right)
\right],
\end{equation}

where $Y$ is the hyper-charge difference. For a charged lepton-neutrino doublet ($Y=-1/2$ for neutrino, $-1$ for charged lepton), which gives $\Delta S \approx 0.0871$.

\begin{figure}[H]
    \centering
    
    \scriptsize
    \begin{tabular}{|c|c|c|c|c|}
        \hline
        $m_{\ell_4}$ (GeV) & $m_{\nu_4}$ (GeV) & $\Delta m$ (GeV) & $\Delta T$ & $\Delta S$ \\
        \hline
        500  & 500  & 0   & 0.00  & 0.053 \\
        500  & 370  & 130 & 0.31  & 0.069 \\
        500  & 100  & 400 & 2.60  & 0.138 \\
        \hline
        750  & 750  & 0   & 0.00  & 0.053 \\
        750  & 620  & 130 & 0.34  & 0.063 \\
        750  & 100  & 650 & 6.60  & 0.160 \\
        \hline
        1000 & 1000 & 0   & 0.00  & 0.053 \\
        1000 & 870  & 130 & 0.26  & 0.060 \\
        1000 & 100  & 900 & 12.40 & 0.175 \\
        \hline
    \end{tabular}
   
    \label{tab:deltaT_deltaS_values}

    \vspace{0.5cm} 

    \includegraphics[width=1.0\linewidth]{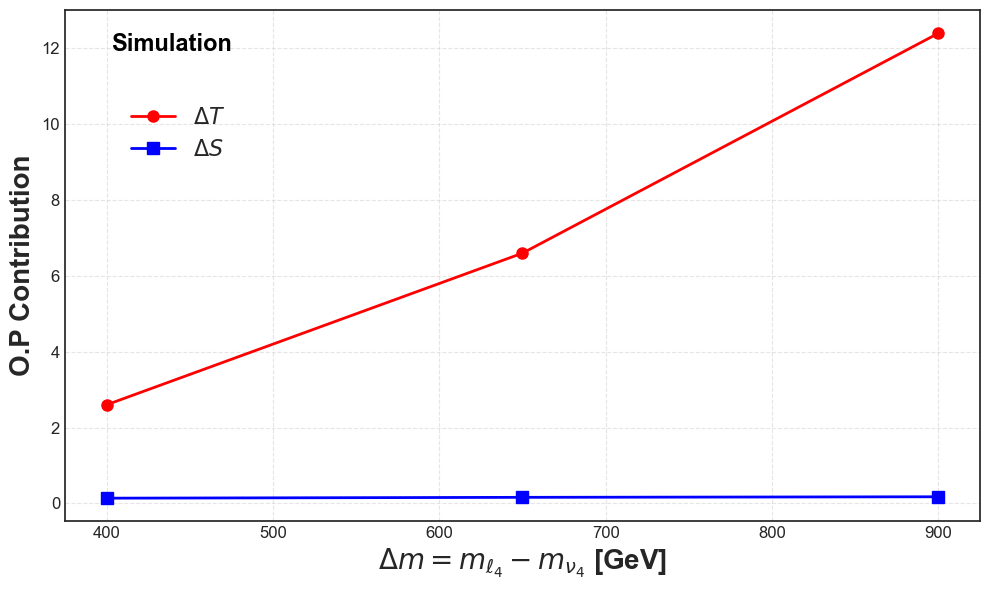}
    \caption{Table shows contribution of S and T parameters for different masses of $\ell_4$ with (i) fixed mass of $\nu_4$ at 100 GeV and (ii) fixed mass splitting of $\Delta m$ = 130 GeV. Mass splitting is chosen to be at the threshold of the limit allowed by electroweak precision tests. Figure shows contributions of $\Delta T$ and $\Delta S$ as a function of mass splitting $\Delta m$ for fixed $m_{\nu_4} = 100 \ \text{GeV}$. }
    \label{fig:oblique_plot}
\end{figure}

Contribution of $T$ parameter grows rapidly with increasing mass splitting. Each additional fermion doublet contributes additively to the oblique parameters $S$ and $T$, where $S$ reflects the overall size of the new sector and $T$ measures its weak isospin breaking. A degenerate heavy generation contributes approximately $1/6\pi$ to S, leading to shifts in the $W$ boson mass and the $Z^0$ polarization asymmetry, as noted by Bertolini and Sirlin. For higher masses of $\ell_4$, neutrino masses have to be large enough to compensate the splitting. The table in Fig.(\ref{fig:oblique_plot}) presents results for different $\ell_4$ and $\nu_4$ masses and their respective contributions to the $T$ and $S$ parameters.

For no contribution to $T$ parameter from mass splitting the mass of lepton and neutrino has to be identical, which is only possible with an off-shell $W$ boson. In this work we do not consider that scenario. These results illustrate that a high mass splitting within the fourth-generation lepton doublet induces a substantial positive $T$ parameter and a moderate positive $S$ parameter, impacting global electroweak fits. These constraints also justify the choice of a Dirac fourth-generation neutrino with mass 100 GeV and a charged lepton mass of 190 GeV, ensuring consistency with direct LEP search limits, while respecting electroweak precision constraints on mass splitting and negligible contributions to precision observables due to suppressed mixing with lighter generations. For heavier leptons, a very heavy neutrino is required to respect the mass splitting and also leads to higher contribution of $S$ and $T$ parameter.

\section{Signal and background cross-sections in  proton scattering}

We perform a simulation preliminary at $\sqrt{s} = 13 \ TeV$. Proton scattering pair produces $\ell_4$, giving  $\ell_4^+ \,(W^+ \nu_4)\;\; \ell_4^- \,(W^- \bar{\nu}_4)$ signature. This process is mediated either by $\gamma, Z$ or $H$. Feynman diagrams of concerned interactions are provided below.

\begin{figure}[H]
  \centering

\begin{minipage}{0.5\textwidth}
  \centering
  \resizebox{\linewidth}{!}{
    \begin{tikzpicture}
      \begin{feynman}
        \vertex (q1) at (-3,1) {q};
        \vertex (q2) at (-3,-1) {$\bar{q}$};
        \vertex (v_in) at (-1.5,0) [dot]{};

        \vertex (v_prop) at (1.2,0) [dot]{};

        \vertex (l4v) at (3.2,1) {$\ell_4$};
        \vertex (l4barv) at (3.2,-1) {$\bar{\ell}_4$};

        \vertex (l4dec) at (4.6,1) [dot]{};
        \vertex (l4dec2) at (4.6,-1) [dot]{};

        \vertex (Wpos) at (6,1.6) {$W^+$};
        \vertex (nu4)  at (6,0.4) {$\nu_4$};
        \vertex (nu4bar) at (6,-0.4) {$\bar{\nu}_4$};
        \vertex (Wneg) at (6,-1.6) {$W^-$};

        \diagram*{
          (q1) -- [fermion] (v_in),
          (q2) -- [anti fermion] (v_in)
        };

        \draw[very thick, decorate, decoration={snake}] (v_in) to[bend left=12] node[above] {$\gamma/Z$} (v_prop);
        \draw[very thick, dashed] (v_in) to[bend right=12] node[below] {H} (v_prop);

        \diagram*{
          (v_prop) -- [fermion] (l4v) -- [fermion] (l4dec),
          (v_prop) -- [anti fermion] (l4barv) -- [anti fermion] (l4dec2),
          (l4dec)  -- [boson] (Wpos),
          (l4dec)  -- [fermion] (nu4),
          (l4dec2) -- [boson] (Wneg),
          (l4dec2) -- [anti fermion] (nu4bar)
        };
      \end{feynman}
    \end{tikzpicture}
  }
  \caption*{}
\end{minipage}

  \caption{Allowed production and decay diagrams for $q\bar{q} \rightarrow\ell_{4}\bar{\ell_{4}}$ with allowed propagators of $\gamma,Z,H$. Final collider signature can be either of type $2\ell+\text{ME}_T$,$4q+\text{ME}_T$ or $2q+\ell+\text{ME}_T$ based on the decay mode of $W$ boson.}
  \label{fig:feynman_l4_production_corrected}
\end{figure}

We define the proton contents in the cross-section calculations (CalcHEP) to be quarks ($u,d,s,t$) and gluons. Though the gluon diagrams are included in sub-processes, cross-section for gluon induced processes is found to be zero. Additionally, processes of the type $q_{i}q{j} \rightarrow \ell_{4}\bar{\ell_{4}}$ also have zero cross-section. We chose the parton distribution function for proton to be \texttt{NNPDF-lo-as-0130-qed (proton)} \cite{LorceMetzPasquiniSchweitzer2025, Belyaev_2013}. Due to two chosen quark generations and gluons, there are totally 17 sub-processes. With increasing mass of $\ell_4$, cross-sections for all mediators reduce significantly. The cross-section has dominant contributions from photon-mediated processes. Higgs-mediated processes have extremely small cross-section, making them practically unavailable at the given collider energy. In this analysis, we have considered only s-channel contributions from all mediators, which allows us to write the E$_{com}$ in terms of the Mandelstam variable ($\sqrt{s}$) \cite{Griffiths2008, Mandelstam1958, HalzenMartin1984,Perkins2000}. The majority contribution from photon mediated processes is due to its propagator structure as a result of masslessness of photon and its coupling to electric charge. $Z$ boson is significantly heavier while Higgs coupling to fermions is governed by fermion mass \cite{Himpsel2015, Himpsel2018, AitchisonHey2004, Grigo2013}.

\begin{table*}[t]
    \centering
    \caption{Total cross-sections for a fourth generation lepton (signal) and backgrounds at various masses and center-of-mass energies in $pp$ scattering for process $pp \rightarrow \ell_4 \bar{\ell}_4\ $.}
    \label{tab:total_xsec}
    \scriptsize   
    \setlength{\tabcolsep}{4pt} 
    \renewcommand{\arraystretch}{1.2} 
    \begin{tabular}{@{}cccccccc@{}}
        \toprule
        & \multicolumn{4}{c}{Signal} & \multicolumn{3}{c}{Background} \\
        \cmidrule(lr){2-5} \cmidrule(lr){6-8}
        $\sqrt{s}$ [TeV] 
        & $190$ GeV/\(c^2\) 
        & $500$ GeV/\(c^2\) 
        & $750$ GeV/\(c^2\) 
        & $1000$ GeV/\(c^2\) 
        & $t\bar{t}$ 
        & $WW$ 
        & $ZZ$ \\
        \midrule
        3  & $4.36 \times 10^{-3}$ & $4.36 \times 10^{-6}$ & $3.19 \times 10^{-8}$ & $2.42 \times 10^{-10}$ & $5.71 \times 10^{0}$ & $6.97 \times 10^{0}$ & $9.06 \times 10^{-1}$ \\
        5  & $1.80 \times 10^{-2}$ & $9.29 \times 10^{-5}$ & $3.84 \times 10^{-6}$ & $1.89 \times 10^{-7}$  & $3.10 \times 10^{1}$ & $1.76 \times 10^{1}$ & $2.39 \times 10^{0}$ \\
        8  & $4.94 \times 10^{-2}$ & $5.79 \times 10^{-4}$ & $5.20 \times 10^{-5}$ & $6.49 \times 10^{-6}$  & $1.16 \times 10^{2}$ & $3.66 \times 10^{1}$ & $5.09 \times 10^{0}$ \\
        10 & $7.48 \times 10^{-2}$ & $1.15 \times 10^{-3}$ & $1.29 \times 10^{-4}$ & $2.08 \times 10^{-5}$  & $2.05 \times 10^{2}$ & $5.03 \times 10^{1}$ & $7.08 \times 10^{0}$ \\
        13 & $1.17 \times 10^{-1}$ & $2.31 \times 10^{-3}$ & $3.18 \times 10^{-4}$ & $6.36 \times 10^{-5}$  & $3.86 \times 10^{2}$ & $7.20 \times 10^{1}$ & $1.03 \times 10^{1}$ \\
        18 & $1.95 \times 10^{-1}$ & $4.89 \times 10^{-3}$ & $8.09 \times 10^{-4}$ & $1.95 \times 10^{-4}$  & $8.06 \times 10^{2}$ & $1.10 \times 10^{2}$ & $1.59 \times 10^{1}$ \\
        20 & $2.28 \times 10^{-1}$ & $6.09 \times 10^{-3}$ & $1.06 \times 10^{-3}$ & $2.68 \times 10^{-4}$  & $1.02 \times 10^{3}$ & $1.26 \times 10^{2}$ & $1.82 \times 10^{1}$ \\
        22 & $2.62 \times 10^{-1}$ & $7.37 \times 10^{-3}$ & $1.34 \times 10^{-3}$ & $3.52 \times 10^{-4}$  & $1.24 \times 10^{3}$ & $1.42 \times 10^{2}$ & $2.06 \times 10^{1}$ \\
        \bottomrule
    \end{tabular}
\end{table*}

\begin{figure*}[t]  
    \centering

    \begin{minipage}{0.48\linewidth}
        \centering
        \includegraphics[width=\linewidth]{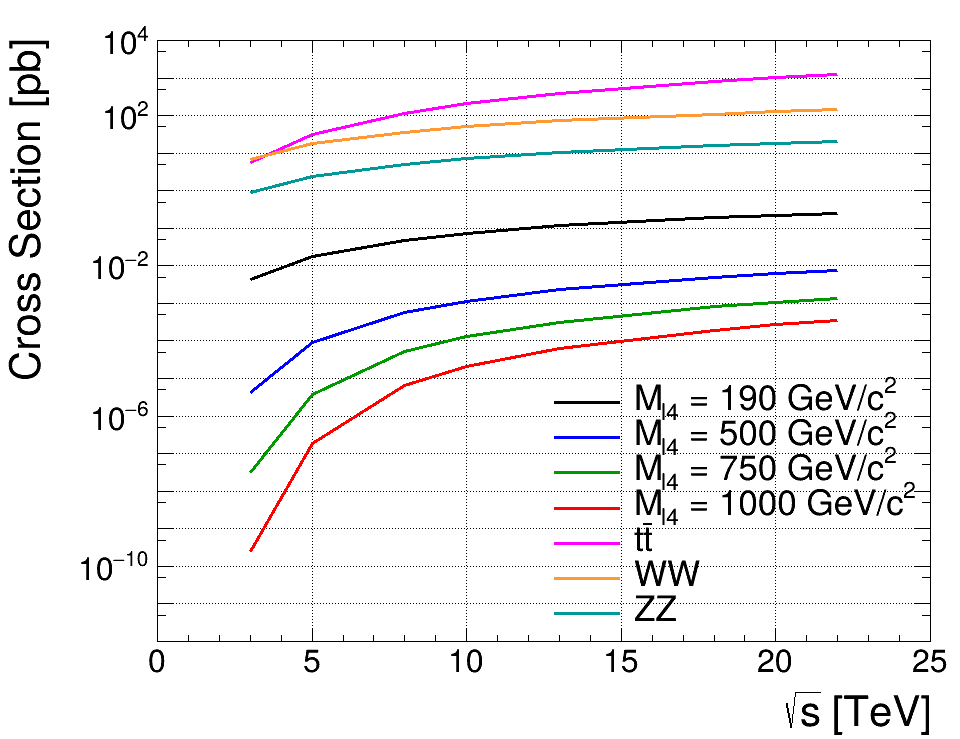}
        \label{cscomA}
    \end{minipage}
    \hfill
    \begin{minipage}{0.48\linewidth}
        \centering
        \includegraphics[width=\linewidth]{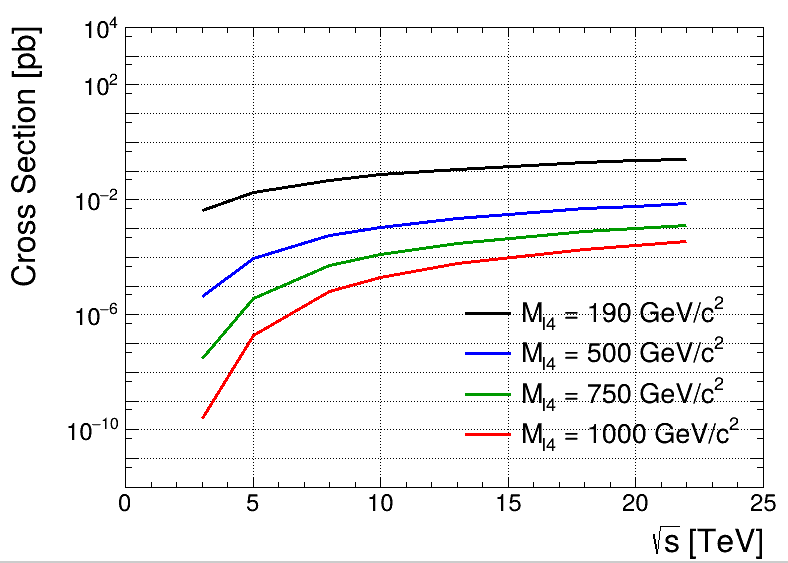}
        \label{fig:fig_b}
    \end{minipage}

    \vspace{0.5cm} 

    \begin{minipage}{0.52\linewidth}
        \centering
        \includegraphics[width=\linewidth]{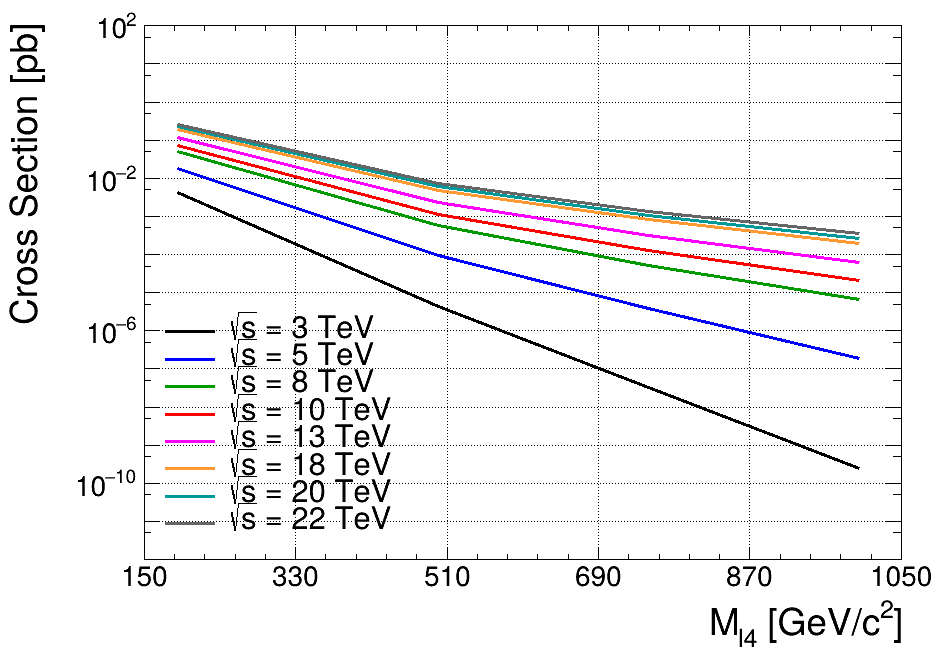}
        \label{fig:fig_c}
    \end{minipage}

    \caption{Variation of cross-sections in $pp$ scattering with mass and centre-of-mass energies. Top row shows signal and background vs ECM, bottom row shows cross-section vs mass.}
    \label{fig:combined_three}\vspace*{-12pt}
\end{figure*}

Energies available to LHC at present times can reach upto 13.6 TeV. In proton colliders, increase in the cross-section with increasing E$_{com}$ is a well-known feature. Hence, for Monte Carlo (MC) simulations we can explore higher E$_{com}$ for studies on probing these New Physics signatures in future colliders like HL-LHC \cite{brüning2025lhcoperationhighluminositylhc}. However, one major obstacle in dealing with high-E$_{com}$ is equivalent scaling of background cross-sections, which can highly contaminate the signal. The impact of scaling the E$_{com}$ on the cross-sections of signal and background events can be seen from Fig.(\ref{fig:combined_three}). Fig.(\ref{fig:combined_three})(top left and right) show characteristic scaling of cross section of the process with increasing E$_{com}$. Fig.(\ref{fig:combined_three})(bottom) shows variation of cross-section with mass at different E$_{com}$. Decreasing cross-section with increasing $\ell_4$ mass is verified with that plot. The trend of increasing cross-sections with increasing E$_{com}$ remains constant across different mass hypotheses of $\ell_4$ as well as background. In general, background cross-sections are orders of magnitude higher than signal and hence one requires an effective cut flow to extract signal. $t\bar{t}$ is the most dominant background. QCD multijets can also fake the signal \cite{PhysRevD.99.114012}.

 We derive cross-sections for QCD background from PYTHIA internal dataset. Probing the lepton BSM signature is highly improbable at lower E$_{com}$ and very heavy $\ell_4$ hypotheses due to very small cross-sections. At high E$_{com}$, the major issue is to suppress background to get a $2\sigma$ significance to either look for possibility discovery or exclude the model at $95 \%$ Confidence Level (C.L.).

\section{Cut flow and signal selection}

For effective analysis of jet and lepton parameters of interest, we apply finite detector resolution and subject all reconstructed final-state objects 
to energy smearing. The energy of a particle, $E_{\text{true}}$, was modified according to a Gaussian distribution with a resolution $\sigma(E)$ as \cite{atlas-cern2013,Delphes2014}:

\begin{align}
E_{\text{smeared}} &= E_{\text{true}} + \Delta E,\\& \nonumber
\quad \Delta E \sim \mathcal{N}(0, \sigma^2(E_{\text{true}}))
\end{align}

where $\mathcal{N}(0, \sigma^2)$ denotes a normal distribution with mean zero and variance $\sigma^2$. Different resolution is used for lepton and hadrons to mimic CMS/ATLAS detector effects. For charged leptons and hadrons (jets), the relative energy resolution was parameterized as \cite{atlas-cern2013},

\begin{equation}
\begin{aligned}
\frac{\sigma(E_{\text{leptons}})}{E_{\text{leptons}}} 
    &= \sqrt{\left(\frac{0.01}{\sqrt{E/\text{GeV}}}\right)^{2} + (0.005)^{2}}, \\
\frac{\sigma(E_{\text{hadrons}})}{E_{\text{hadrons}}} 
    &= \sqrt{\left(\frac{0.10}{\sqrt{E/\text{GeV}}}\right)^{2} + (0.020)^{2}} .
\end{aligned}
\end{equation}

The expression is composed of energy dependent noise which is the first term in square root and the constant term which accounts for calibration and instrumental errors. Then we use smeared energies to recalculate the momentum components:

\begin{align}
p_x^{\text{smeared}} &= p_x^{\text{true}} \frac{E_{\text{smeared}}}{E_{\text{true}}}, \\
p_y^{\text{smeared}} &= p_y^{\text{true}} \frac{E_{\text{smeared}}}{E_{\text{true}}}, \\
p_z^{\text{smeared}} &= p_z^{\text{true}} \frac{E_{\text{smeared}}}{E_{\text{true}}}.
\end{align}

This procedure ensures that kinematic variables such as transverse momentum ($p_T$), pseudorapidity ($\eta$), and missing transverse energy are subject to realistic detector effects while preserving the direction of the original particle momentum. Followed by this, we employ a standard cut flow.

\begin{figure}[H]
\centering
\begin{tikzpicture}[
  box/.style={draw, thick, rounded corners, inner sep=2mm, text width=6cm, align=center, font=\small},
  arrow/.style={-{Latex[length=2.5mm]}, thick}
  ]

\node[box, fill=gray!20] (cuts) {Selection Cuts};

\node[box, below=0.5cm of cuts, fill=red!10] (baseline) {
$|\eta_\ell|<2.5,~|\eta_j|<4.5$\\
$p_T^\ell>40~\mathrm{GeV},~p_T^j>20~\mathrm{GeV},~p_T^{\mathrm{miss}}>30~\mathrm{GeV}$\\

\textbf{(Baseline selection cuts)}};

\node[box, below=0.5cm of baseline, fill=blue!10] (bkg) {
$n_\ell \ge 1,~n_j=2$,
$b$-veto: $\epsilon_b=90\%,~c\to b=10\%,~\text{light}\to b=1\%$\\
$\Delta\phi>1.5\ rad$
\textbf{($t\bar{t},ZZ,QCD$ suppression)}};

\node[box, below=0.5cm of bkg, fill=green!10] (signal) {
$-2.5 \le \eta_\ell \le -0.05,~ -4.5 \le \eta_j \le -0.5$
\textbf{(Signal isolation)}};

\draw[arrow] (cuts.south) -- (baseline.north);
\draw[arrow] (baseline.south) -- (bkg.north);
\draw[arrow] (bkg.south) -- (signal.north);

\end{tikzpicture}
\caption{Cutflow diagram and motivation for each cut. Standard baseline cuts combined with tight cut on jet and lepton multiplicity along with effective b-veto, significantly reduces $t\bar{t}$ and QCD backgrounds.}
\label{fig:cutflow_diagram}
\end{figure}
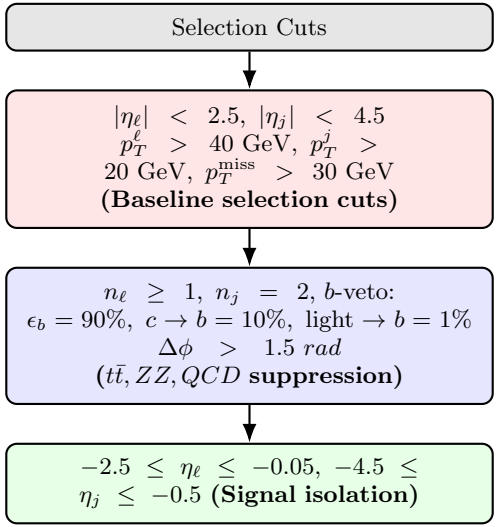

$\eta$ cuts for lepton and jets are standard CMS preliminary cuts applied in MC studies. Jet and lepton $p_T$ cuts are moderate. These cuts are motivated from observed $p_T$ peaks from MC simulation. Based on the decay mode of each $W$ boson from signal, the type of signatures that signal can give are $2\ell+\text{ME}_T, 1\ell+2j+\text{ME}_T,4j+\text{ME}_T$. Based on this, lepton and jet multiplicity cuts are very helpful in the analysis.Lepton multiplicity cut along with lepton $p_T$ essentially cuts down significantly on QCD background primarily since high $p_T$ leptons are produced in QCD with low probability while signal can produce upto 2 leptons in case of leptonic decay of both $W$s. Though we employ shape study of jet multiplicity, jet multiplicity cut significantly reduced $t\bar{t}$ background since the background produces more than 2 jets in many events and signal can produce single lepton, dijet. Though this cut aggressively cuts down on the case of signal events producing 4 jets through completely hadronic decays of $W's$, surviving signal fraction, relative to surviving background fraction,still gives a good signal selection efficiency (refer Table \ref{tab:cutflow_comparison}). The multiplicity cut coupled with b-veto almost nullifies the $t\bar{t}$ background thus significantly improving local significance of the signal.  To suppress backgrounds with heavy-flavor jets, a $b$-jet veto was applied at the analysis level. 
Jets were reconstructed for $R=0.4$, 
using all final-state hadrons and photons as inputs after detector smearing. 
The heavy-flavor content of each reconstructed jet was determined by tracing its constituents 
back to the Pythia event record. If any $B$-hadron was found among the constituents of a jet, 
the jet was tagged as originating from a $b$-quark. In cases where no $B$-hadron was present but a 
$C$-hadron was identified, the jet was assigned as charm-flavored. 
If neither was found, the jet was classified as a light-flavor jet. 
To account for realistic detector performance, $b$-tagging was applied probabilistically: 
$b$-jets were tagged with an efficiency of $\epsilon_{b} = 90\%$, while charm and light-flavor jets 
were mistagged with rates of $\epsilon_{c} = 10\%$ and $\epsilon_{l} = 1\%$, respectively. 
An event was vetoed if at least one reconstructed jet was tagged as a $b$-jet under this procedure. 
This $b$-veto substantially reduces the contribution from top-quark pair production and other 
backgrounds with real $b$-jets, while retaining a high signal efficiency. Relatively relaxed $\text{ME}_T$ cut is applied since a high $\text{ME}_T$ cut distorts the signal invariant mass peak although high $\text{ME}_T$ cut can effectively reduce both $t\bar{t}$ and QCD backgrounds. Also due to particular nature of signatures from signal and backgrounds, we use specific strategies of invariant mass reconstruction. These strategies are explained in detail in later sections.

For this analysis, a total of $n_{\text{event}} = 100{,}000$ events were generated for signal and background $t\bar{t}, WW$ and $ZZ$. For QCD background, we use $\hat{p}_{T}$
binned distributions with five different bins, namely; $100 \ GeV<\hat{p}_{T}<200 \ GeV, 200 \ GeV<\hat{p}_{T}<300 \ GeV, 300 \ GeV<\hat{p}_{T}<400 \ GeV $ and $400 \ GeV<\hat{p}_{T}<500 \ GeV$ and for each bin generate $n_{\text{event}} = 100{,}000$. For each bin, the generator computes the corresponding production cross-section,
denoted by $\sigma_{\text{bin}}$, through an internal integration of the perturbative
QCD matrix elements over the parton distribution functions subject to the
phase-space constraints. Explicitly, the bin cross-section is given by
\begin{equation}
\sigma_{\text{bin}}
=
\int_{p_{T,\text{min}}}^{p_{T,\text{max}}}
\frac{d\sigma}{dp_{T}} \, dp_{T},
\end{equation}
with
\begin{align}
\frac{d\sigma}{dp_{T}}&=
\sum_{i,j} \int dx_{1} \, dx_{2} \;
f_{i}(x_{1}, Q^{2}) \, \\ \nonumber &f_{j}(x_{2}, Q^{2}) \,
\frac{d\hat{\sigma}_{ij \to kl}}{dp_{T}} \, ,
\end{align}
where $f_{i,j}(x, Q^{2})$ are the parton distribution functions evaluated at momentum
fraction $x$ and scale $Q^{2}$, and $d\hat{\sigma}_{ij \to kl}$ denotes the
partonic-level scattering cross-section for incoming partons $i,j$.
The value of $\sigma_{\text{bin}}$ returned by the generator corresponds to the inclusive
cross-section within the chosen $\hat{p}_{T}$ interval and is subsequently used to
normalize the MC sample to particular luminosity.
Jets were reconstructed using the anti-$k_{T}$ algorithm from PYTHIA 8.313 with a clustering radius of $R = 0.4$. 
The event yields were normalized to an integrated luminosity of $\mathcal{L} = 138~\text{fb}^{-1}$.

\begin{table}[H]
\centering
\scriptsize
\resizebox{\columnwidth}{!}{
\renewcommand{\arraystretch}{0.95}
\setlength{\tabcolsep}{3pt}
\begin{tabular}{lcccccc}
\hline\hline
\textbf{Sample} & \multicolumn{2}{c}{$N_{\ell}\!\geq\!1$} & \multicolumn{2}{c}{$b$-veto} & \multicolumn{2}{c}{$N_{j}\!=\!2$} \\
                & Off & On & Off & On & Off & On \\
\hline
QCD\_100--200 & 0.0031 & 0.0032 & 0.0008 & 0.0011 & 0.0003 & 0.0003 \\
QCD\_200--300 & 0.0122 & 0.0124 & 0.0038 & 0.0040 & 0.0007 & 0.0005 \\
QCD\_300--400 & 0.0218 & 0.0211 & 0.0076 & 0.0075 & 0.0011 & 0.0010 \\
QCD\_400--500 & 0.0304 & 0.0298 & 0.0109 & 0.0107 & 0.0015 & 0.0014 \\
QCD total     & 0.0169 & 0.0161 & 0.0058 & 0.0058 & 0.0009 & 0.0008 \\
Signal        & 0.2007 & 0.2022 & 0.1839 & 0.1845 & 0.0602 & 0.0597 \\
$WW$          & 0.1525 & 0.1553 & 0.1415 & 0.1442 & 0.0468 & 0.0473 \\
$ZZ$          & 0.0992 & 0.0999 & 0.0779 & 0.0788 & 0.0256 & 0.0259 \\
$t\bar t$     & 0.2398 & 0.2392 & 0.0099 & 0.0102 & 0.0016 & 0.0017 \\
\hline
              & \multicolumn{2}{c}{$\text{ME}_T$ cut} & \multicolumn{2}{c}{Final sel.} & \multicolumn{2}{c}{} \\
              & Off & On & Off & On & & \\
\hline
QCD\_100--200 & 0.00002 & 0.00004 & 0.00002 & 0.00004 & & \\
QCD\_200--300 & 0.00028 & 0.00023 & 0.00028 & 0.00023 & & \\
QCD\_300--400 & 0.00040 & 0.00053 & 0.00040 & 0.00053 & & \\
QCD\_400--500 & 0.00066 & 0.00071 & 0.00066 & 0.00071 & & \\
QCD total     & 0.00034 & 0.00038 & 0.00034 & 0.00038 & & \\
Signal        & 0.0482  & 0.0481  & 0.0482  & 0.0481  & & \\
$WW$          & 0.0275  & 0.0271  & 0.0275  & 0.0271  & & \\
$ZZ$          & 0.0109  & 0.0108  & 0.0109  & 0.0108  & & \\
$t\bar t$     & 0.0013  & 0.0014  & 0.0013  & 0.0014  & & \\
\hline\hline
\end{tabular}
}
\caption{Signal selection efficiency post each cut and post whole cutflow are reported in the table above. Final selection efficiencies are reported in a normalized fashion for a total generated event pool of $1e5$ events.'off' column reports numbers for ISR/FSR-Off and 'on' column reports selection for ISR/FSR-On MC samples. Final efficiences are found to be $\varepsilon_{\text{sig}}^{\text{Off}} = 0.0482$, $\varepsilon_{\text{sig}}^{\text{On}} = 0.0481$, 
$\varepsilon_{t\bar{t}}^{\text{Off}} = 0.0013$, $\varepsilon_{t\bar{t}}^{\text{On}} = 0.0014$, 
$\varepsilon_{WW}^{\text{Off}} = 0.0275$, $\varepsilon_{WW}^{\text{On}} = 0.0271$, 
$\varepsilon_{ZZ}^{\text{Off}} = 0.0109$, $\varepsilon_{ZZ}^{\text{On}} = 0.0108$.}
\label{tab:cutflow_comparison}
\end{table}


The anti-$k_t$ algorithm is widely preferred because it produces jets with perfectly conical, circular boundaries that are unaffected by soft radiation. This greatly reduces sensitivity to underlying event and pileup contamination \cite{CMS2020_PileupMitigation, BertaMasettiMillerSpousta2019}. 

\begin{figure}[H]
    \centering
    \includegraphics[width=0.9\linewidth]{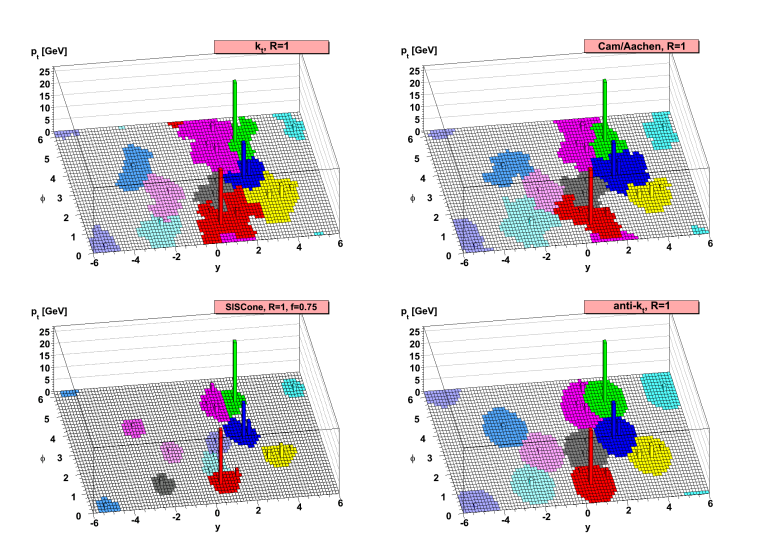}
    \caption{Comparison of clustering performed by different clustering algorithms on soft ‘ghost’ particles to demonstrate efficiency of each clustering algorithm. SIScone and anti-$k_t$ algorithms have less rugged shapes for soft and hard jets making them suitable for jet clustering applications with high E$_{com}$ \cite{CacciariSalamSoyez2008}.}
    \label{fig:anti_kt_graphic}
\end{figure}

The figure shows application of different clustering algorithms to $\approx 10^4$ random soft ‘ghost’ particles demonstrated in \cite{CacciariSalamSoyez2008}. Uneven clustering boundaries in $k_t$ and Achen algorithms are results of the sensitiveness of these algorithms to the set of ghosts involved. For the SIScone algorithm, single particle jets are regular on contrary to composite jets. Anti-$k_t$ algorithm on the other hand produces pronounced circular shape for hard jets and soft jets have varied complex shapes. This demonstrates the effectiveness of these algorithms in jet clustering. Since we are working high E$_{com}$, we expect hard jets to be produced and hence it is justified to use the anti-$k_t$ algorithm for effective clustering of the same.\\

\section{Jet and lepton kinematics}

Considering the decay mode of $\ell_4$, we reconstruct $W$ boson from the dijet invariant mass for signal. Similarly, invariant mass of other background can also be reconstructed.

\begin{figure}[H]
    \centering

    \includegraphics[width=\linewidth]{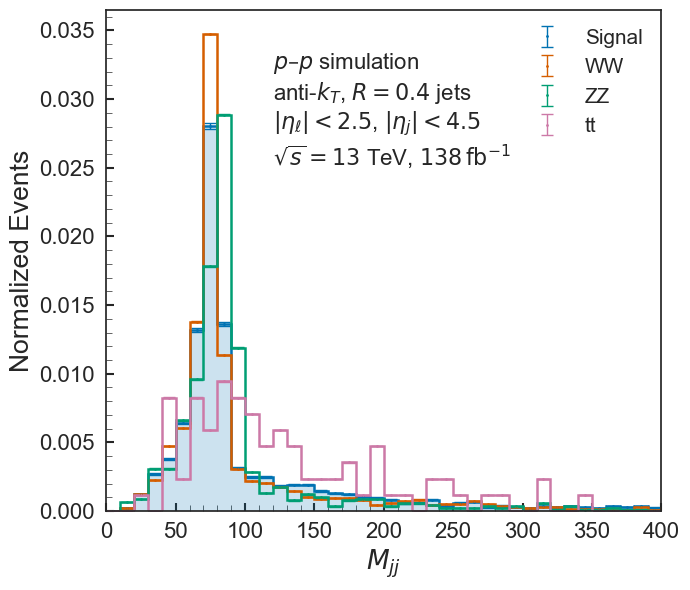}
    \caption{Dijet invariant mass distributions for proton scattering with ISR/FSR on. Dijet mass for signal peaks at around $W$ mass.}
    \label{fig:dijet_inv_mass_190_ISR_FSR}
\end{figure}

As can be seen from plots above, dijet invariant mass for signal peaks in the range 72--82 GeV consistent with $W$ boson mass. Similarly, for $WW$ and $ZZ$ background, peaks in the range 72--82 GeV and 82--92 GeV are consistent with true masses. Due to b-veto applied for $t\bar{t}$ suppression, scaled  $t\bar{t}$ distribution peaks much lower than other backgrounds. The dijet mass peak for  $t\bar{t}$ with ISR/FSR effects is not very clear but is in the range of 82--92 GeV consistent with reconstructed mass of $W$ boson from the $t(W^+b)\bar{t}(W^-\bar{b})$ decay. ISR/FSR effects make the distribution broader. With no ISR/FSR we get cleaner peaks including $t\bar{t}$ background.

In this setup, signal and $t\bar{t}$ background are very heavy compared to other background. $t\bar{t}$ prefers central production of jets and lepton. Signal jets and lepton production prefers $\eta<0$ making it an important parameter to distringhuish signal from heavily contaminating $t\bar{t}$ background. On the contrary, light backgrounds $WW$ and $ZZ$ are boosted and prefer moderately forward-background production of jets and leptons. 

There is no preferred $\phi$ direction for jets and leptons. Hence in the $\eta-\phi$ plane,backgrounds can mimic and contaminate signal mainly in the negative $\eta$ region. An $\eta$ cut can still efficiently isolate the signal. This is evident from Fig.({\ref{fig:kinematics_190_ISR_FSR_on}})

In terms of jet multiplicity also, signal is not entirely distinguishable from backgrounds. Varying clustering radius with anti-$k_t$can have significant impact on jet multiplicity distributions however for this study we are not concerned with those systematic effects. Hence keeping the clustering radius $R=$ 0.4, we study jet multiplicities.

\begin{figure}[H]
    \centering
    \includegraphics[width=\linewidth]{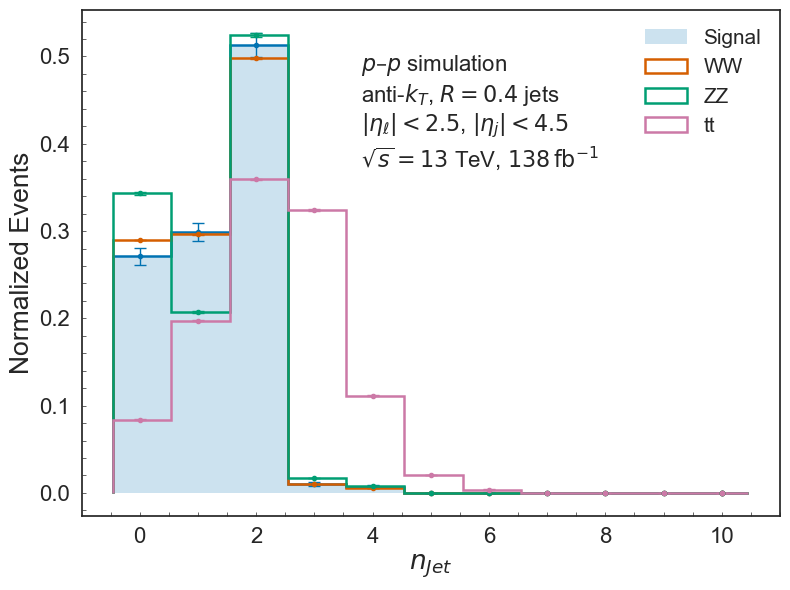}
    \vspace{4pt}
    \includegraphics[width=\linewidth]{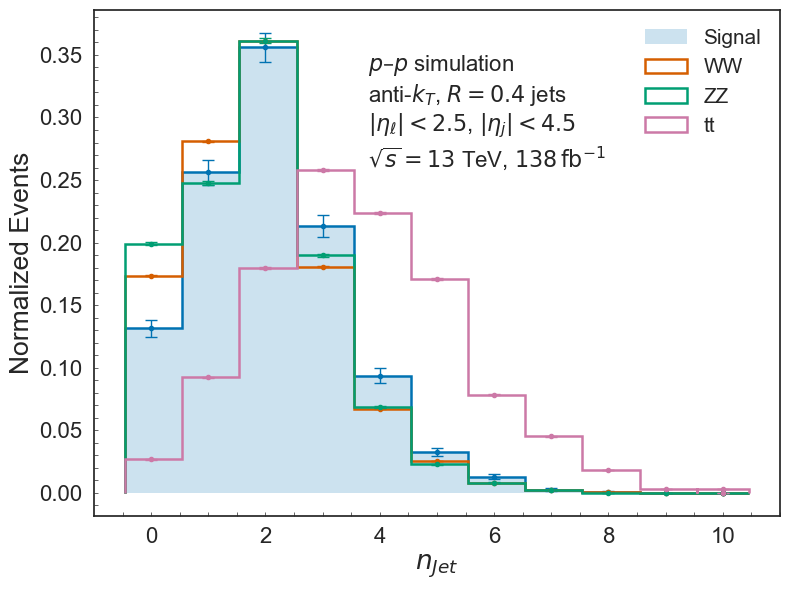}
    \caption{Jet multiplicity distributions in proton scattering with jet clustering radius $R=0.4$: ISR/FSR off (top) and ISR/FSR on (bottom). Signal events prefer 2 jets which are consistent with hadronic decay of one $W$ and leptonic of other.}
    \label{fig:jet_mult_190_ISR_FSR}
\end{figure}

\begin{figure*}[t]
    \centering
    \begin{minipage}{0.4\linewidth}
        \centering
        \includegraphics[width=\linewidth]{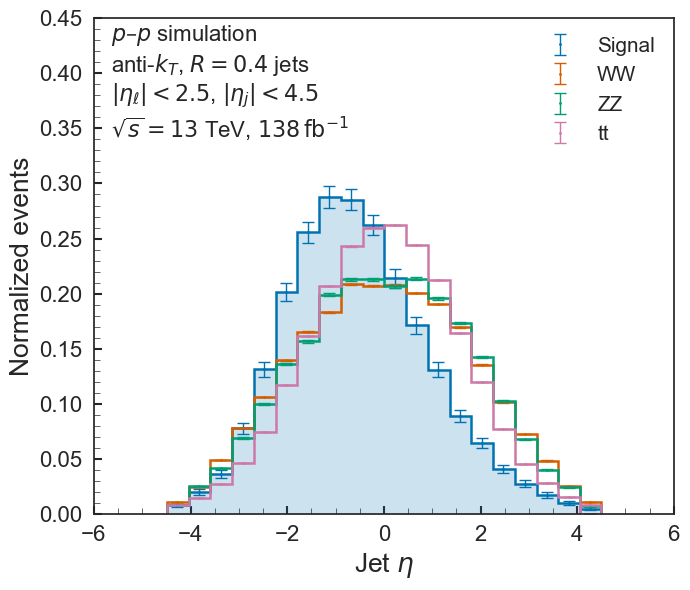}
    \end{minipage}
    \begin{minipage}{0.4\linewidth}
        \centering
        \includegraphics[width=\linewidth]{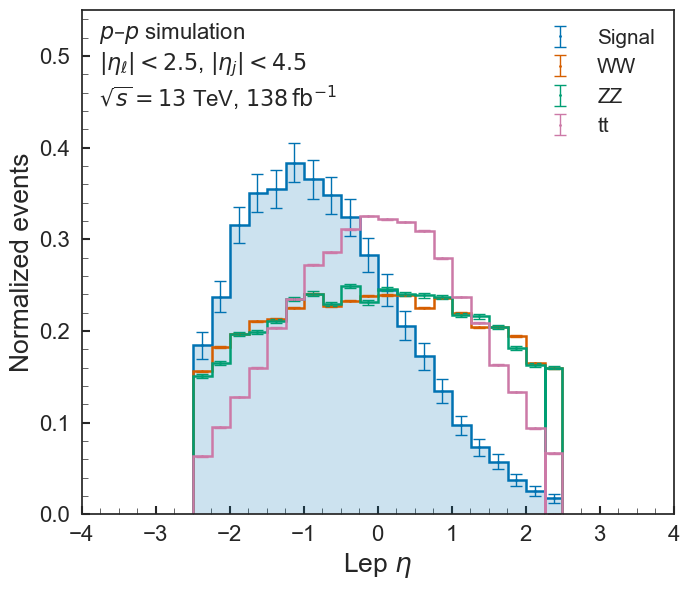}
    \end{minipage}

    \vspace{0.5cm} 

        \begin{minipage}{0.4\linewidth}
        \centering
        \includegraphics[width=\linewidth]{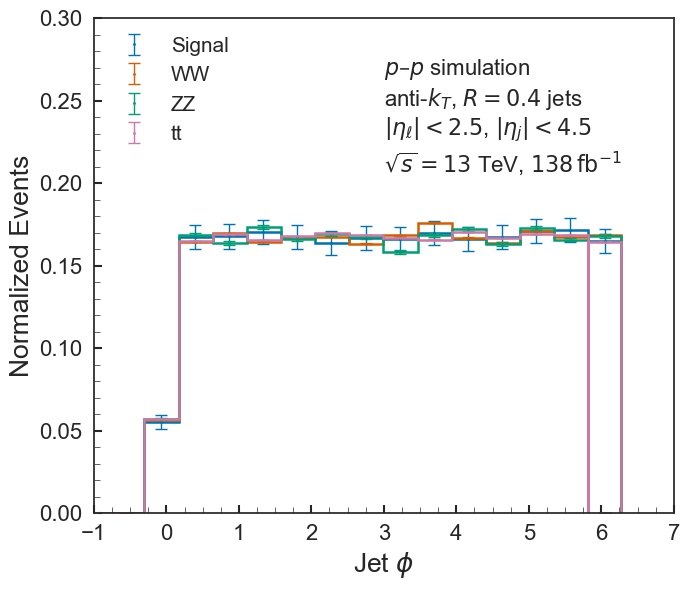}
    \end{minipage}
    \begin{minipage}{0.4\linewidth}
        \centering
        \includegraphics[width=\linewidth]{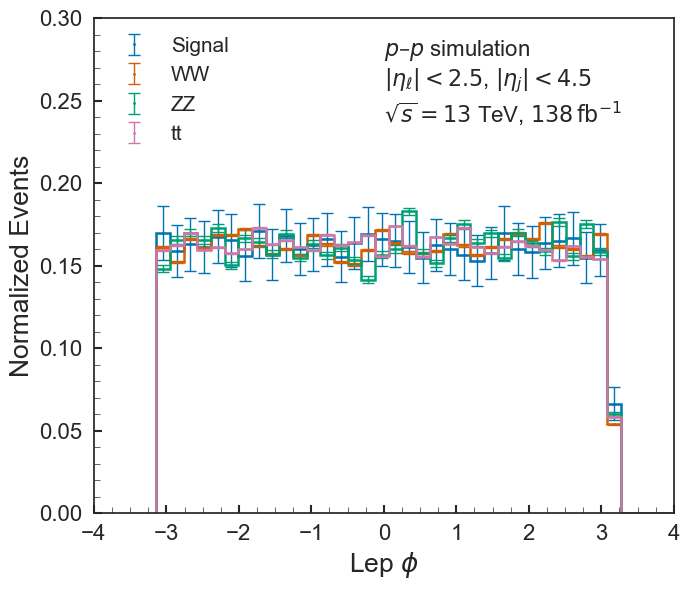}
    \end{minipage}\hfill

    \caption{$\eta-\phi$ distributions for proton scattering with ISR/FSR on. Jet $\eta$ (top left),lepton $\eta$ (top right),jet $\phi$ (bottom left), and lepton $\phi$ (bottom left). Signal and backgrounds are distinguishable in the $\eta$ direction in $\eta-\phi$ plane for both jets and leptons.}
    \label{fig:kinematics_190_ISR_FSR_on}
\end{figure*}

\begin{figure}[H]
   \begin{minipage}{\linewidth}
        \centering
        \includegraphics[width=\linewidth]{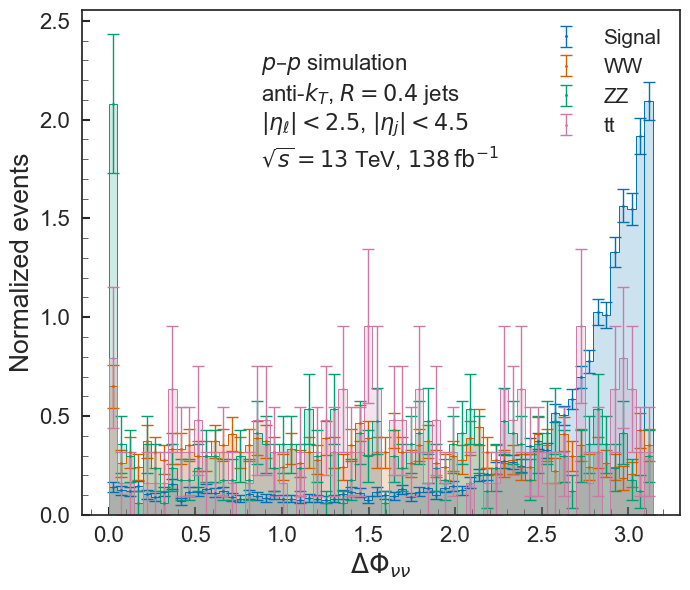}

        \label{fig:delta_phi_with_met_sig_bg}
    \end{minipage}

    \caption{Comparison of distributions of $\Delta \phi$ between BSM signal neutrinos and SM background neutrinos, with $\text{ME}_T$ cut. This is an important distinguishing feature for signal.}
    \label{fig:delta_phi_neutrinos_grid}\vspace*{-12pt}
\end{figure}

For signal, jets arise from $W$ decays. Leptonic decays of $W$ will not give any jets but hadronic decays will give 2 jets. Since two W's are produced in the decay of two $\ell_4$s, maximum possible jets are 4 and minimum possible jets are 0. Each of the $W$ can decay either leptonically or hadronically. Both W's decaying hadronically gives 4 jets and both W's decaying leptonically gives 0 jets. One $W$ decaying hadronically and one $W$ decaying leptonically gives 2 jets. ISR/FSR effect significantly impact the distribution and gives many more gluon induced and hadronization based jets. Irrespective of ISR/FSR effects, signal has peak at $n_{jet}=2$ and tail reaches upto 4 jets. Sources of these jets are already mentioned. As a sanity check, one can observe that for signal, maximum of 4 jets are present with no ISR/FSR effects as expected. Background also have similar peaks. $ZZ$ and $WW$ have decay channels which can give 4 jets for hadronic and 2 jets for semi-hadronic decays. This can be easily verified from the multiplicity plot. Tails of both of these background reach upto 4 jets with peak at 2 jets. For $t\bar{t}$ background, one can have upto 6 jets due to the $bW(jj)$ decay channel and hence the tail for $t\bar{t}$ reaches upto 6 jets. With ISR/FSR effects, tail of both signal and backgrounds extend further as expected. Peak of only $t\bar{t}$ shifts, but other backgrounds and signal peaks remain intact.\\
Due to the $W+\text{ME}_T$ decay mode the heavy BSM lepton, missing transverse momentum is an essential parameter to infer presence of fourth generation neutrino. Plots show consistent peaks for backgrounds. Since leptonic decay of $W$ and invisible deacy of $Z$, can produce neutrinos one can expect peaks at approximately half the parent mass for these backgrounds in rest frame in mass limit. Due to partial cancellation of momenta, peak shifts to a slightly lower value. This can be verified for backgrounds. In case of signal, heavy back to back neutrinos are produced as evident from plots above.The peak at $\delta\phi \approx \pi$, in fact indicates that neutrinos are produced back to back.\\
Without $\text{ME}_T$ cut, this distribution is much smoother in contrast to $\text{ME}_T$ cut. Due to $\text{ME}_T$ cut, fraction of events are rejected. $\text{ME}_T$ cut noticably reduces background and fake soft neutrino peaks. In the transverse plane,this $\Delta\phi$ distribution this leads to reduction in the net $\text{ME}_T$ thus downshifting the signal $\text{ME}_T$ peak.\\
$\delta\phi$ is a very important distinguishing feature between signal and background. As can be observed from Figs.(\ref{fig:delta_phi_neutrinos_grid}); for signal $\Delta\phi \approx \pi$ is a very clear distinguishing peak from background which are mostly flat with peaks at $\Delta\phi \approx 0$. These peaks are false peaks arising from soft neutrinos from ISR/FSR and hadronization. This indicates strong preference of signal for back to back production while for background there is momenta cancellation in the plane leading to flatter distributions.\\
The available phase space for these neutrinos corresponds to $\Delta m=$ 90 GeV [refer Table (\ref{tab:l4widths_updated})] for signal. Hence tail of the signal distribution can reach upto those values. The reminiscent parts of the distribution are a result of boost effects. Distorted distribution for $t\bar{t}$ background is a result of veto applied. In conclusion, $\delta\Phi$ cut can effectively separate signal from background increasing the observed local significance. Though our cutflow is derived from basic principles and does not include an advanced cut on $\phi$ parameter,we study the effect of $\delta\phi$ cut on global and local significance in the 180--300 GeV signal window.\\
The reconstruction of the $\ell_4$ mass is facilitated by $\text{ME}_T$ values. During reconstruction it is important to take into account the massive neutrino and the longitudinal momentum of the neutrino.Signal was reconstructed using $W+\text{ME}_T$. The $W$ boson candidate was reconstructed from dijet invariant mass of two leading jets, where the dijet invariant mass $M_{jj}$ was computed as;

\begin{equation}
M_{jj} = \sqrt{(E_{j1}+E_{j2})^{2} - (\vec{p}_{j1}+\vec{p}_{j2})^{2}}.
\end{equation}

Missing transverse momentum, $p_{T}^{\,miss}$ is used as a proxy for neutrino, and the reconstructed $\ell_{4}$ mass was then obtained from the four-vector sum of the $W$ candidate and the $\text{ME}_T$ for heavy $\nu_4$.

 In the case of $WW$ production, the invariant mass was calculated from the dijet system, with a mass window of 500 GeV. This mass window is chosen to account for the tail of $W$ distribution contaminating the signal in the mass window 180--300 GeV. For $ZZ$ production, opposite-charge, same-flavor lepton pairs were combined to reconstruct $Z \to \ell^{+}\ell^{-}$ candidates, with the dilepton invariant mass computed from the corresponding lepton four-momenta. In the $t\bar{t}$ background, the invariant mass of the top quark was reconstructed by combining three jet candidates from hadronic decay of $W$ and b ($W(q\bar{q})b)$. For the QCD multijet background, the dijet invariant mass was directly used from internal PYTHIA calculations. At hadron colliders, the unknown longitudinal momentum of the partonic system can prevent full invariant mass reconstruction. So one may use the transverse mass ($M_T$) as a proxy observable sensitive to the mass of $\ell_4$. $M_T$ is defined as,

\begin{equation}
M_T
=
\sqrt{
(E_T^W + E_T^{\nu_4})^2
-
|\vec{p}_T^W + \vec{p}_T^{\nu_4}|^2
},
\end{equation}

In the context of this study, MC simulation allows for a complete reconstruction. But in experimental analyses, $M_T$ can serve as a valid observable to interpret signal.

\begin{figure}[t]
    \centering

    \begin{minipage}{\linewidth}
        \centering
        \includegraphics[width=\linewidth]{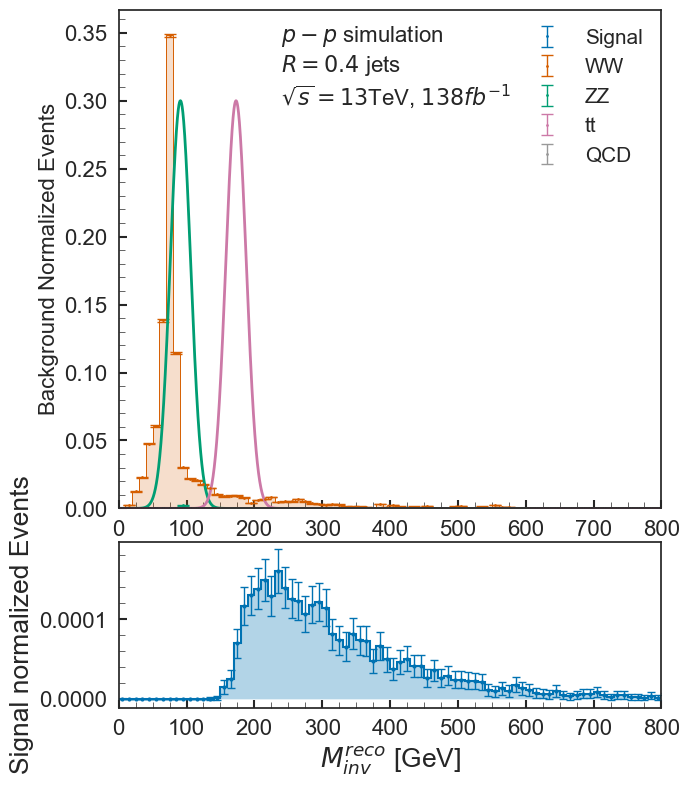}
    \end{minipage}
    
    \vspace{10pt} 

    \begin{minipage}{\linewidth}
        \centering
        \includegraphics[width=\linewidth]{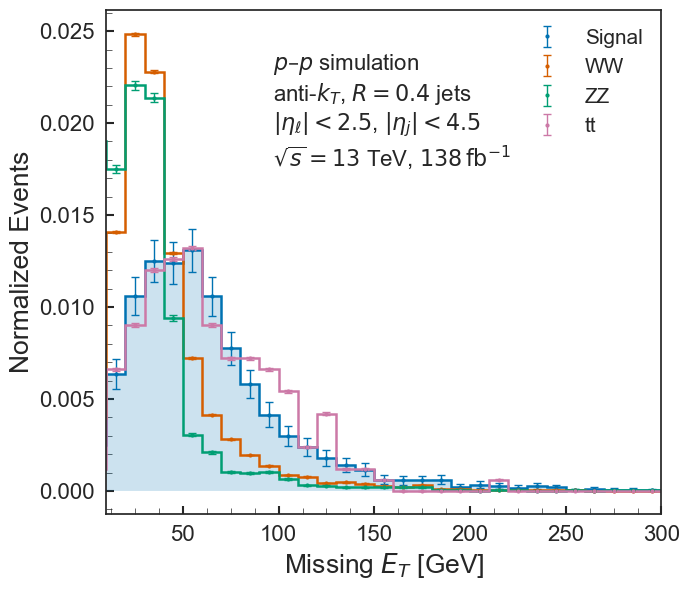}
    \end{minipage}

    \caption{Reconstruction of the fourth-generation lepton using $2j+\text{ME}_T$ signature (Top). Cutflow removes majority of ZZ,$t\bar{t}$ and QCD background.Hypothetical Gaussians (green:$ZZ$,red:$t\bar{t}$), are drawn to represent respective background invariant mass peaks. Both of these backgrounds can significantly contaminate signal region. Missing transverse energy ($p_T^{\mathrm{miss}}$) distribution for signal and background with ISR/FSR (Bottom). True peak downshifts mainly due to preference for back to back production of $\nu_{4}$.}
    \label{fig:combined_plots}\vspace*{-12pt}
\end{figure}

This mass window is chosen pertaining to the spread of the signal distribution. Since there is a noticeable spread in the signal around the true mass range of 180--220 GeV, we choose the given mass window to keep majority of the signal though signal tail is lost. In this analysis, $\delta\phi$ cut is applied to reject events if the angular separation value is lesser than given value.

\begin{figure*}[t]
    \centering
    \begin{tabular}{cccc}
        \includegraphics[width=0.45\textwidth]{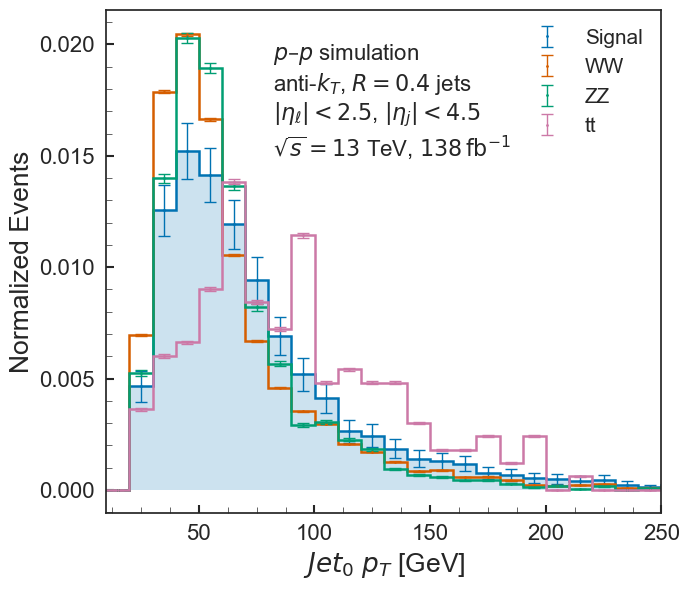} &
        \includegraphics[width=0.45\textwidth]{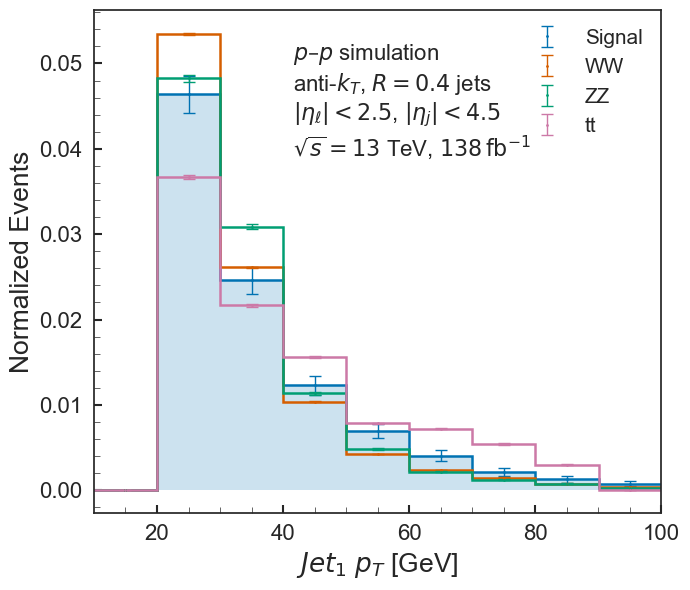}  \\
        \\[6pt]

        \includegraphics[width=0.45\textwidth]{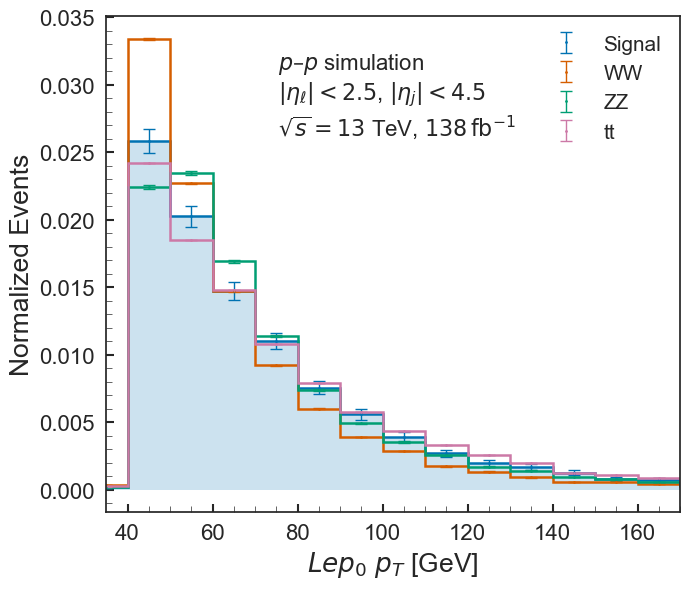} &
        \includegraphics[width=0.45\textwidth]{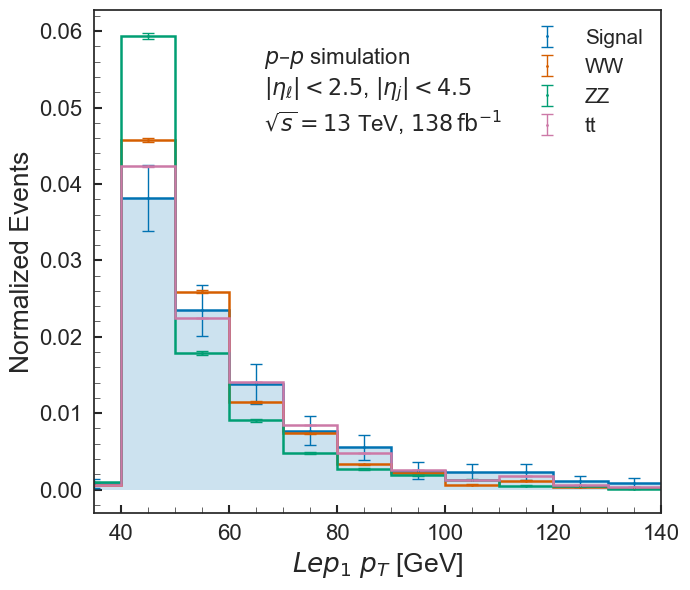}  \\
    \end{tabular}
    \caption{Comparison of transverse momentum ($p_T$) distributions for leading jet (top left), subleading jet (top right), leading lepton (bottom left) and subleading lepton (bottom right).}
    \label{fig:jet_lep_pt_comparison}\vspace*{-12pt}
\end{figure*}

Pin pointing sources of jets is a straightforward task. Detailed explanation of jet sources is already provided in the jet multiplicity section. For signal, jets come from the $W$ boson. In the $W$ boson rest frame, ideally one expects jet $p_T$ to peak at $\approx 40$ GeV for leading jet. However in the parent rest frame, $W$ acquires lesser momentum leading to even softer jets. Hence the leading jet $p_T$ for signal is in the expected range. The hard tail of $t\bar{t}$ background is expected due to high $p_T$ b jets. Very soft jet events are rejected due to jet $p_T$ cut of 20 GeV. ISR/FSR effect slightly up-shift the jet $p_T$ peak alongside slightly broader tails. Similarly, in case of leptons, for signal, leptonic decay of $W$ boson can produce lepton $p_T$ in the range 30--50 GeV range, consistent with MC results.

\section{Discovery potential}

\begin{figure}[H]
    \centering
    \includegraphics[width=1.0\linewidth]{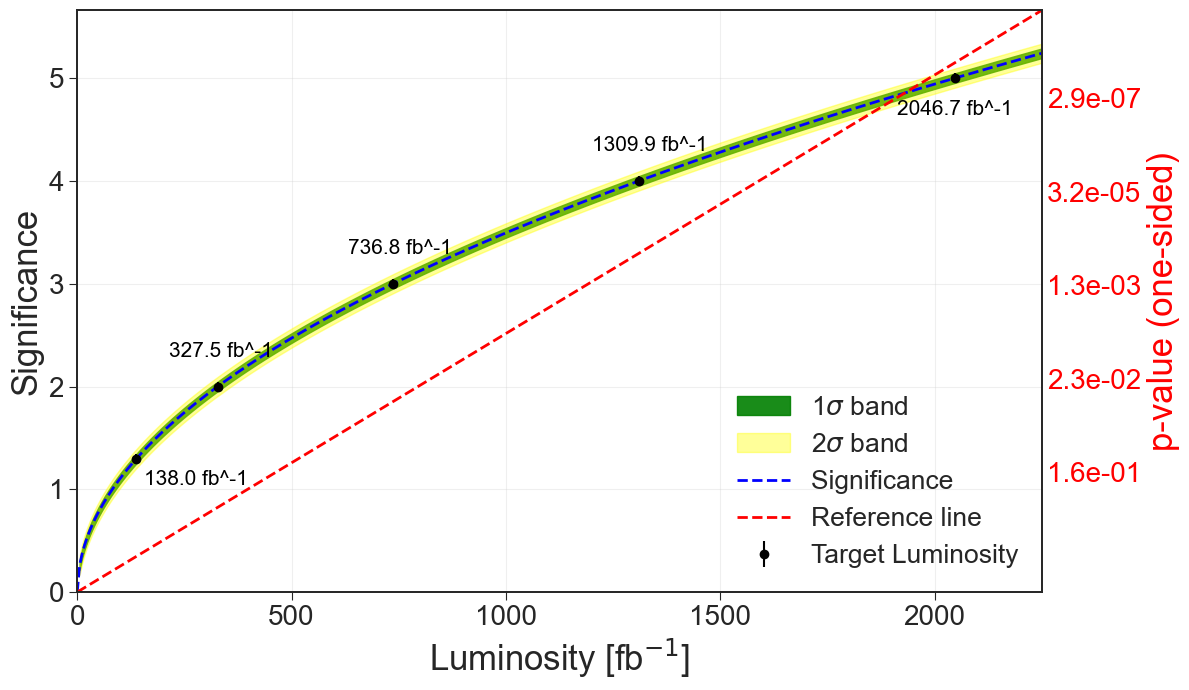}
    \caption{Significance of signal discovery as a function of luminosity for masses of leptons under consideration with $1\sigma$ and $2\sigma$ bands. It would require nearly 200 times the present luminosity to reach a $5\sigma$ discovery with the current cutflow and signal selection.}
    \label{fig:sigvlumi}
\end{figure}

Discovery potential can be demonstrated with mass and luminosity spectrum. In Fig.(\ref{fig:sigvlumi}), we demonstrate global discovery significance curve for chosen BSM lepton and neutrino of masses of $190\ GeV$ and $100\ GeV$ respectively. At 138 fb$^-1$, global significance is very small at $1.30 \pm 0.065(stat)$ and to reach significance of $5\sigma$ it would require very high luminosity of about 2000 fb$^-1$. This is not a realistically achievable standard in near future. Hence, a realistic analysis goal would be to calculate the significance in the signal region of 180--300 GeV. $\delta\phi$ cut slightly improves upon the local significance. Global significance reaches a maximum value of $1.47 \pm 0.073(stat)$ for a small $\delta\phi$ cut of $0.5\ rad$. Local significance reaches upto $2.76 \pm 0.138(stat)$ for $\delta\phi > 1.5\ rad$.

 There is a slight ever increase in the significance due to this cut. Standard cutflow already achieves a local significance of $2.40 \pm 0.120(stat)$. At, higher angular separation cut, significance (local \& global) drops inspite of larger signal population in that cut region. Since $\delta\phi$ cut succeeds moderate $\text{ME}_T$ cut, some events are already rejected at the $\text{ME}_T$ level. Higher $\text{ME}_T$ cut would keep majority of the back to back neutrino signal events.

Though signal populates higher angular seperation region, substantial number of background events also populate that region as seen from figure 16. Hence rejecting the $\delta\phi$ tail for signal on an average reduces the significance as background distribution somewhat maintain isotropic angular separation distributions. 
\begin{figure*}[t]
    \centering

    \includegraphics[width=0.48\linewidth]{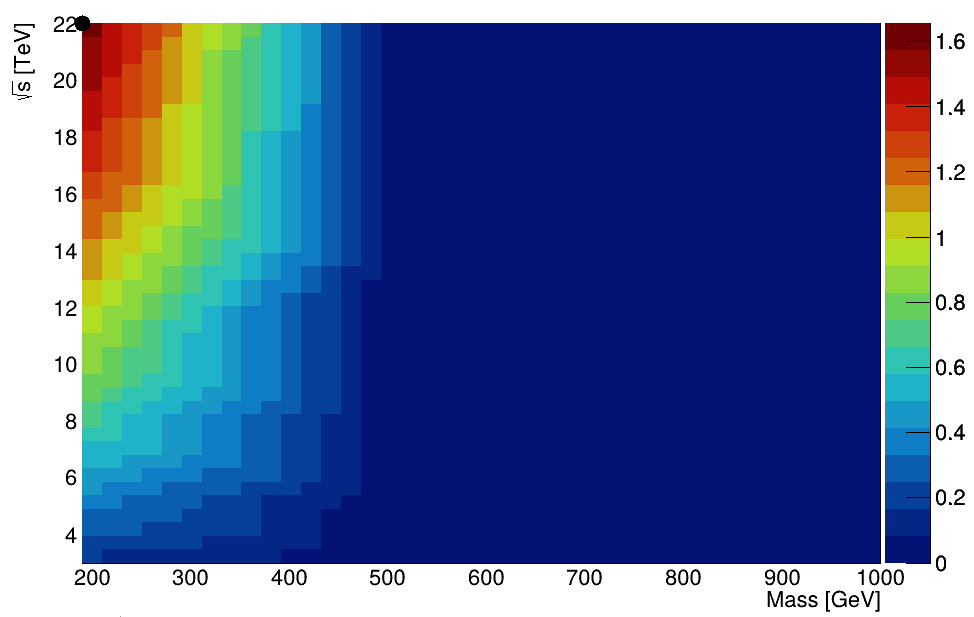}
    \hfill
    \includegraphics[width=0.48\linewidth]{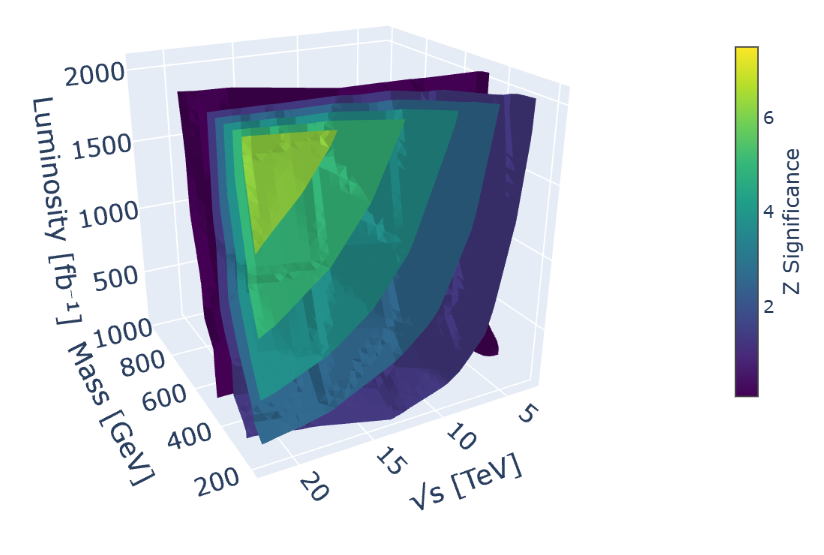}

    \vspace{0.5cm}

    \includegraphics[width=0.48\linewidth]{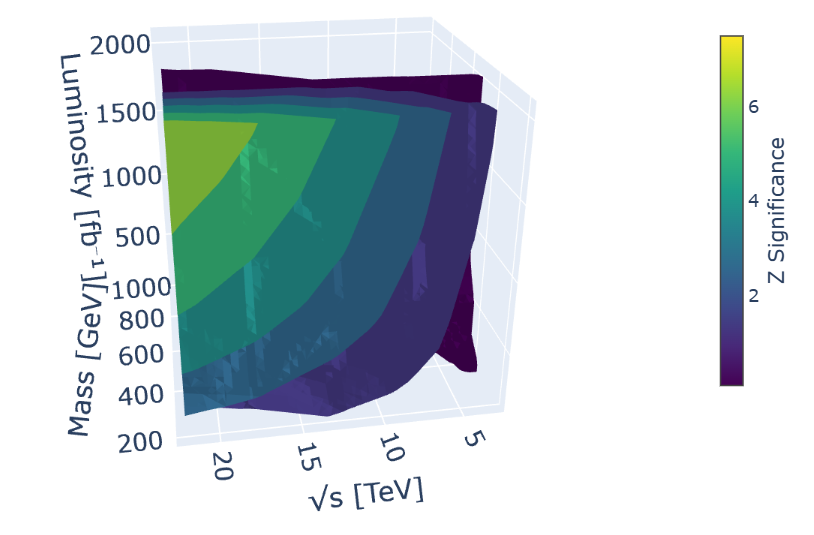}
    \hfill
    \includegraphics[width=0.48\linewidth]{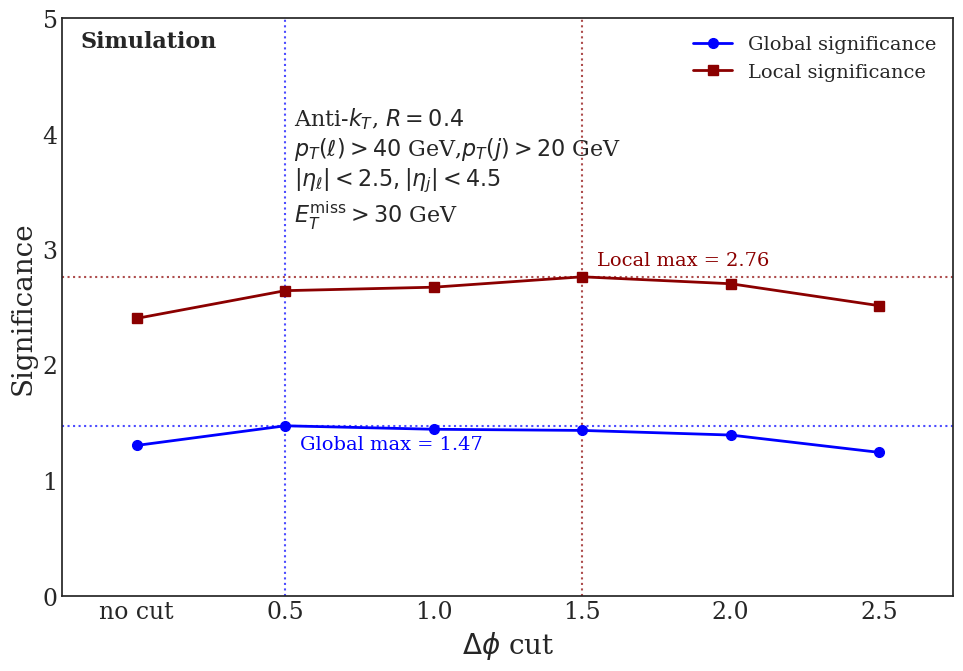}

    \caption{Visualization of discovery significance: three plots (top left, top right and bottom left),show variation across mass and center-of-mass energy planes, with and without luminosity dependence. In the $\sqrt{s}-mass$ plane, significance rapidly drops for high mass hypotheses of $\ell_4$. Higher luminosities are promising for probing these BSM signatures. The fourth plot (bottom right) shows the dependence of significance on the $\delta\phi$ cut, which is motivated by the back-to-back nature of $\nu_4$ production and shows a noticeable improvement over global and local significance values.}
    \label{fig:significance_grid}
\end{figure*}
\\Significance plots are for signal and background selection efficiencies of $\varepsilon_{\text{sig}} = 0.0481$, $\varepsilon_{t\bar{t}} = 0.0014$, $\varepsilon_{WW} = 0.0271$, and $\varepsilon_{ZZ} = 0.0108$. Signal was observed to have a non-symmetric distribution for $\eta_{\ell}\ \&\ \eta_{j}$. $\eta_{\ell}$ peaks at $\approx-1$ and $\eta_{j}$ peaks at $\approx -1.5$. On isolating $-2.5\leq\eta_{\ell}\leq-0.05$, global significance improves to $1.55\pm 0.092(stat)$ and local significance improves upto $3.12\pm0.202(stat)$. On further adding $-0.5<\eta_{j}<-4.5$ cut, global significance slightly drops to $1.46 \pm0.068(stat)$ but local significance improves to $3.33 \pm 0.241(stat)$ proving the efficiency of $\eta$ cuts in the analysis. We report baseline values for our analysis. Further improvement on these values is possible with precision tuning of cuts. 

\section{Conclusion}
In this work we run MC simulation for fourth generation lepton model for a reference lepton mass of $190\ $GeV with the mass splitting of $90\ $GeV in $p-p$ simulation at 13 TeV E$_{com}$. Constraints on fermion masses due to LEP and EWPO are respected and we demonstrate how mass splitting contributes significantly to T parameter for large splitting, making those scenarios improbable. Cross-sections of the process increase significantly at higher E$_{com}$ and can be probed at higher energies and luminosity at LHC, HL-LHC and other future colliders. Very heavy leptons of the order $10^3\ $GeV have rapidly falling cross-section making it improbable to probe them with current luminosity and energies. We have also demonstrated effective cut flow, including advanced cuts on $\delta\phi_{\nu_{4}}$ and $\eta_{\ell},\eta_{j}$, giving us global significance of $1.46 \pm0.068(stat)$ and local significance of $3.33 \pm 0.241(stat)$. The local excess observed is at $99.9\%\text{ C.L}$ and the global excess observed is at $92.7\%\text{ C.L}$. The angular separation between signal neutrinos from the only available $\ell_4^+(W^+\bar{\nu_4})\ell_4^-(W^-\nu_4)$ channel strongly prefer back to back production and hence an angular separation cut further enhances the significance. Detector level b-veto applied to significant $t\bar{t}$ background with optimized tagging efficiencies for b jets and mistag rates, performs well on improving the significance. The jet and lepton parameters of interest also show expected distributions. Discovery significance rapidly drops at higher $\ell_4$ masses and lower center of mass energies. This strongly favors $50\ GeV\leq m_{\ell_4}\leq400\ GeV$ scenarios for fixed $m_{\nu_{4}}=100\ GeV$. Higher luminosities at present colliders and future upgrades, are promising to probe BSM lepton sector. With the present present cutflow, we predict that a luminosity of $\approx 2e3\ fb^{-1}$ is required to claim discovery of $\ell_4$. For a moderate mass $\ell_4$, at about $15-20\ TeV$ center of mass energy, around $300-400\ fb^-1$ is sufficient to reach $3\sigma$.

\nocite{*}

\bibliography{apssamp}

\end{document}